\documentclass[10pt, doublecolumn]{IEEEtran}
\usepackage{amsmath}
\usepackage{amssymb}
\usepackage{cite}
\usepackage{algorithmic}
\usepackage{algorithm}
\usepackage{array}
\usepackage{graphicx}
\usepackage{balance}
\usepackage{setspace}
\graphicspath{{fig/}}
\usepackage{xcolor}
\allowdisplaybreaks[4]

\title{Time-Scale-Adaptable Spectrum Sharing for Hybrid Satellite-Terrestrial Networks}
\author{\IEEEauthorblockN{Yanmin Wang, Wei Feng, \emph{Senior Member, IEEE}, Yunfei Chen, \emph{Fellow, IEEE}, Yongxu Zhu, \emph{Senior Member, IEEE}, Shidong Zhou, \emph{Member, IEEE}, and Cheng-Xiang Wang, \emph{Fellow, IEEE}}
	\thanks{
	Yanmin Wang is with the School of Information Engineering, Minzu University of China, Beijing 100081, China~(email: wangyanmin@muc.edu.cn).
	
	Wei Feng (corresponding author) and Shidong Zhou are with the Department of Electronic Engineering, Tsinghua University, Beijing 100084,
	China~(email: zhousd@tsinghua.edu.cn, fengwei@tsinghua.edu.cn). 
	Wei Feng is also with State Key Laboratory of Space Network and Communications, Tsinghua University, Beijing 100084, China.
	
	Yunfei Chen is with Department of Engineering, University of Durham, Durham DH1 3LE, U.K. (email: yunfei.chen@durham.ac.uk).
	
	Yongxu Zhu and Cheng-Xiang Wang are with National Mobile Communications Research Laboratory, School of Information Science and Engineering, Southeast University, Nanjing 210096, China, and also with the Purple Mountain Laboratories, Nanjing 211111, China (email: yongxu.zhu@seu.edu.cn, chxwang@seu.edu.cn).
	}
}

\begin{document}
    \maketitle

%\begin{spacing}{1.5}
\begin{spacing}{1.0}	
	\begin{abstract}
	Cooperation between satellite and terrestrial wireless networks promises
    great potential in meeting fast-growing demands for ubiquitous communications coverage. 
    To tackle spectrum scarcity, spectrum sharing is studied for a hybrid satellite-terrestrial network 
    where satellite links share the same group of time-slotted subcarriers with terrestrial links opportunistically.
    In particular, 
    with coarse network-wide time synchronization, 
    a time-scale-adaptable spectrum sharing framework is proposed
    based on a satellite-terrestrial cooperation time scale that can be flexibly adjusted according to practical requirements.
    For generality, it is assumed that both full and partial frequency reuse could be adopted among the base stations (BSs)
    and satellite selection is supported when multiple satellites are available.
    Relying on only statistical channel state information (CSI),
    joint link scheduling and power control are explored to maximize the average sum rate of the network
    while ensuring quality of service (QoS) for users.
    To solve the complicated mixed integer programming (MIP) problem,
    a low-complexity spectrum sharing scheme is presented based on link-feature-sketching-aided hierarchical link clustering
    and Monte-Carlo-and-successive-approximation-aided transmit power optimization. 
	Simulation results demonstrate that by link feature sketching, 
	diversity of the links brought by the spatial distribution of the users could be well utilized.
	The proposed scheme promises a significant performance gain even under strict inter-link interference constraints.
		
	\end{abstract}
	
	\begin{IEEEkeywords}
		\emph{Channel state information, hybrid satellite-terrestrial network, link feature, spectrum sharing, time scale} 	
	\end{IEEEkeywords}
\end{spacing}
	
	\IEEEpeerreviewmaketitle
	
%\begin{spacing}{1.53}
\begin{spacing}{1.0}	
\section{Introduction}
Hybrid satellite-terrestrial networks are widely acknowledged as a promising solution for ubiquitous communication coverage
in the sixth-generation (6G) era~\cite{Feng_2025, Feng_2024, SatCom_integrate_book, SatCom_book, r_JSAC_2021}.
To tackle the problem of spectrum scarcity,
spectrum sharing between the satellite and terrestrial components has attracted
a lot of attention~\cite{10678835,r_tccn_2015,r_twc_2017_2,r_JSTSP_2019,r_twc_2022, WangChinacom2018, r_twc_2017, r_twc_2019, Chinacom_2025}.
In practical applications,
spectrum is usually utilized in terms of time-frequency resource blocks (RBs) 
consisting of subcarriers and time slots~\cite{SatCom_integrate_book, SatCom_book, r_JSAC_2021, 10678835}.
Accordingly, 
spectrum sharing could be opportunistically carried out both in the frequency domain over subcarriers and in the time domain over time slots,
so as to fully exploit the diversity of links and achieve 
higher spectrum efficiency~\cite{SatCom_integrate_book, SatCom_book, r_JSAC_2021, 10678835, Chinacom_2025}. 

For satellite-terrestrial spectrum sharing in the time domain, synchronization between links
is a fundamental prerequisite~\cite{SatCom_integrate_book, SatCom_book, r_JSAC_2021, 10678835, Chinacom_2025}.
However, due to significant difference between link delays in the satellite and terrestrial components,
network-wide fine time synchronization, e.g., over ms-level time slots, is rather challenging~\cite{SatCom_book, 10678835}.
As a result, spectrum sharing in the time domain over time slots
is more difficult than that in the frequency domain over subcarriers 
for a hybrid network~\cite{10678835,r_tccn_2015,r_twc_2017_2,r_JSTSP_2019,r_twc_2022, WangChinacom2018, r_twc_2017, r_twc_2019}.
With the idea of circumventing the challenge of network-wide fine time synchronization,
a fine-over-coarse scheme was proposed for the hybrid satellite-UAV-terrestrial maritime network based on link scheduling
in our previous work in~\cite{10678835}.
By link clustering and grouped time slice allocation for link clusters,
satellite-terrestrial synchronization just needed to be achieved coarsely at a time scale much larger 
than an ms-level time slot~\cite{10678835}.
With acceptable performance loss,
spectrum sharing relying on coarse network-wide time synchronization tends to be preferred in practical applications, 
for low link orchestrating complexity and signaling overhead~\cite{SatCom_integrate_book,10678835}.

While our previous work in~\cite{10678835} started exploration on
time-domain coordination between satellite and terrestrial links for spectrum sharing, 
most of existing works only focused on the frequency domain~\cite{r_tccn_2015, r_twc_2017_2, r_JSTSP_2019, r_twc_2022, WangChinacom2018, r_twc_2017, r_twc_2019, Chinacom_2025}.
It results in an inability to fully exploit the time-domain diversity of links in satellite-terrestrial networks.
In this paper, joint time-frequency spectrum sharing with coarse network-wide time synchronization
is explored for a hybrid satellite-terrestrial network.
Different from~\cite{10678835},  joint link scheduling and power control are implemented in this work.
Satellite selection is also considered to adapt to the case when multiple satellites are available in the satellite component.
Link feature sketching in perspectives of both link scheduling and power control is utilized 
to improve the implementation efficiency of spectrum sharing.
Besides, to achieve appropriate trade-offs between performance and complexity,
it is assumed that the time scale for satellite-terrestrial cooperation can be adjusted flexibly according to practical requirements.
For generality, 
both full and partial frequency reuse among base stations (BSs) are supported in the terrestrial component.

\subsection{Related Works}
In previous research, investigation on satellite-terrestrial spectrum sharing mainly focused on the frequency
domain~\cite{r_tccn_2015, r_twc_2017_2, r_JSTSP_2019, r_twc_2022, WangChinacom2018, r_twc_2017, r_twc_2019, Chinacom_2025}.
With instantaneous channel state information (CSI) of the links,
frequency-domain spectrum sharing was studied for hybrid satellite, UAV, and terrestrial networks 
in~\cite{r_tccn_2015,r_twc_2017_2,r_JSTSP_2019,r_twc_2022}.
In~\cite{r_tccn_2015}, 
allocation of resources, including the beam, power, and channel, was optimized for cognitive satellite links 
under constraints on interference caused to incumbent terrestrial networks.
In~\cite{r_twc_2017_2}, outage probability was derived for a hybrid satellite-terrestrial system
where multiple terrestrial transmitter-receiver pairs cooperate with a primary satellite network for dynamic spectrum access.
In~\cite{r_JSTSP_2019}, a joint beamforming and power allocation scheme was proposed to maximize
the sum rate of satellite-terrestrial integrated networks, where the satellite multicast downlinks and the terrestrial downlinks
serving a group of non-orthogonal multiple access users share the same mmWave frequency band.
With the presence of a primary satellite-receiver link,
multiple UAVs with aerial stations and a terrestrial BS were deployed to support smart vehicles,
and user association, power optimization, and trajectory control were jointly optimized in~\cite{r_twc_2022}.
No time-domain optimization was considered for spectrum sharing in~\cite{r_tccn_2015, r_twc_2017_2, r_JSTSP_2019}.
Although optimization over time slots was involved in~\cite{r_twc_2022}, it was only for UAV and terrestrial links,
and the primary satellite-receiver link was supposed to occupy the frequency band statically all the time.

While~\cite{r_tccn_2015,r_twc_2017_2,r_JSTSP_2019,r_twc_2022} 
have demonstrated the promising performance of satellite-terrestrial spectrum sharing in hybrid networks,
it is difficult to obtain instantaneous CSI of all links due to its fast variation 
as well as the large delay of satellite links~\cite{r_feng_1, r_feng_2}.
To circumvent the challenge, 
spectrum sharing based on statistical CSI 
was addressed for hybrid satellite-terrestrial networks in~\cite{WangChinacom2018, r_twc_2017, r_twc_2019, Chinacom_2025}.
With only information of path loss and shadowing obtained from a pre-constructed radio map,
power allocation and user scheduling schemes were proposed for a hybrid network 
with satellite-terrestrial spectrum sharing in~\cite{WangChinacom2018}.
With the assumption that only the mean and variance of channel gains are available for interference links,
resource scheduling was discussed in~\cite{r_twc_2017} for mobile wireless and meteorological satellite services
sharing the same frequency band.
With statistical or instantaneous interference constraints imposed by primary terrestrial links,
energy efficient power allocation was investigated for cognitive satellite-terrestrial networks in~\cite{r_twc_2019}.
With the target of maximizing both the number of served terrestrial MTs and the average sum transmission rate, 
a double-target spectrum sharing scheme is proposed for hybrid satellite-terrestrial networks
in~\cite{Chinacom_2025} based on large-scale CSI. 
Just like~\cite{r_tccn_2015,r_twc_2017_2,r_JSTSP_2019,r_twc_2022},
transmission characteristics of the satellite link/links were assumed to stay invariant 
in~\cite{WangChinacom2018, r_twc_2017, r_twc_2019, Chinacom_2025},
and no optimization across time was considered for the satellite component.

When time-domain optimization is to be carried out for satellite-terrestrial cooperation,
the basic time scale needs to be selected appropriately according to specific network scenarios~\cite{IEEE_Network_2024,IoT_2024}. 
In~\cite{IEEE_Network_2024},
an autonomous resource management architecture is proposed for 6G satellite-terrestrial integrated networks,
and an artificial-intelligence-centric three-level resource management closed-loop is designed.
It could be inferred that the time scale for the implementation of resource management is a fundamental issue.
To tackle with differentiated time granularity resource scheduling requirements,
a two-timescale hierarchical bandwidth allocation algorithm 
is proposed based on multi-agent deep reinforcement learning for multi-beam satellite networks in~\cite{IoT_2024}.

\subsection{Main Contributions}
In this paper, we focus on statistical-CSI-based spectrum sharing for a hybrid satellite-terrestrial network,
where satellite links share the same group of time-slotted subcarriers with terrestrial links opportunistically.
Unlike~\cite{WangChinacom2018, r_twc_2017, r_twc_2019, Chinacom_2025},
our work considers joint time-frequency coordination among satellite and terrestrial links. 
The main contributions are as follows.
\begin{itemize}
	\item A time-scale-adaptable spectrum sharing framework is proposed based on 
	coarse network-wide time synchronization.
	To maximize the average sum rate while ensuring quality of service (QoS) for users,
	joint link scheduling and power control are implemented based on statistical CSI of the links.
	Satellite selection is also supported to adapt to the case
	when multiple satellites are available in the satellite component.
	For generality, either full or partial frequency reuse can be adopted among BSs in the terrestrial component.	

	\item After a complicated mixed integer programming (MIP) problem is formulated,
	link features in perspectives of both link scheduling and power control are designed following the divide-and-conquer principle.
	With the aid of link feature sketching,
	link scheduling is decoupled with power control for implementation efficiency.
	A hierarchical link clustering scheme is then proposed for link scheduling under the time-scale-adaptable spectrum sharing framework.
		
	\item A low-complexity spectrum sharing scheme is proposed 	based on link-feature-sketching-aided hierarchical link clustering
	and Monte-Carlo-and-successive-approximation-aided transmit power optimization.
	Simulation results demonstrate that by link feature sketching, 
	link diversity brought by the spatial distribution of the users could be well utilized
	for opportunistic spectrum sharing.
	The proposed scheme promises a significant performance gain even under strict inter-link interference constraints.

\end{itemize}

\section{System Model and Problem Formulation}
\subsection{System Model}
\begin{figure} [t]
	\centering
	\includegraphics[width=8cm]{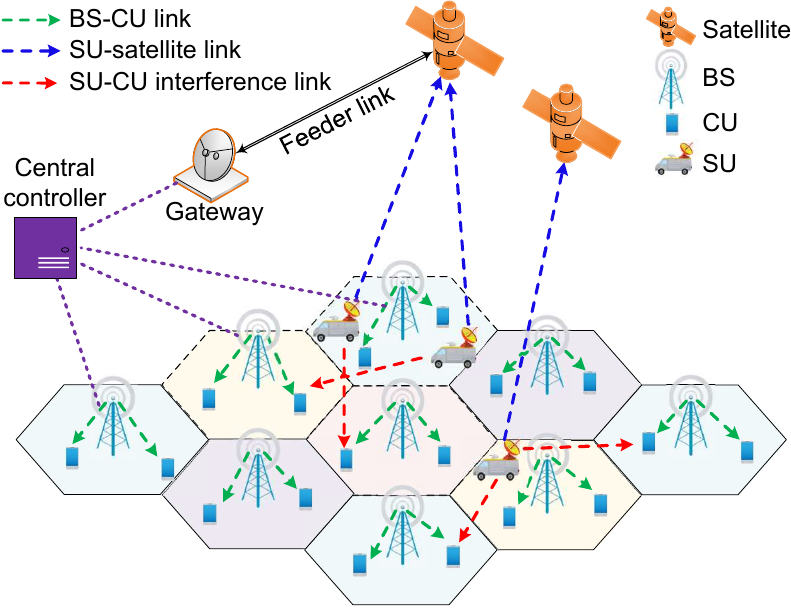}
	\caption{Illustration of a hybrid satellite-terrestrial network.}
	\label{fig_network}
\end{figure}

A hybrid satellite-terrestrial network is shown in Fig.~\ref{fig_network}.
In the network, there are $J$ satellites and $M$ ground BSs.
The satellites are deployed at an altitude of $\mathcal{H}_{sat}$ and serve $N_s$ satellite users (SUs) in the uplink 
via SU-satellite links.
For each SU, only one satellite is selected to serve it at each specific time, 
although multiple satellites could be available simultaneously.  
Each of the BSs serves $N_c$ cellular users (CUs) in the downlink via BS-CU links.
The SUs and CUs either stay still or move randomly within the coverage area of the network.
To utilize the spectrum efficiently, the SU-satellite links share the same frequency band with the BS-CU links,
under the coordination of a central controller.

Suppose $K$ subcarriers, each with a bandwidth of $B$, are available for the hybrid network in
the serving time with a duration of $T$.
The BSs utilize the subcarriers based on a frequency reuse factor of $F \in \mathbb{Z}^+$, 
with $1\leq F \leq K$~\cite{SatCom_integrate_book}. 
Note that $F=1$ corresponds to full frequency reuse among the BSs.
In Fig.~\ref{fig_network}, $F=4$ is shown as an example, and cells with the same color will be allocated with the same group of subcarriers.
Without loss of generality, it is assumed that
\begin{subequations}\label{eq0_0}
	\begin{align}
	& M = I_{cl} F, \\
	& K = K' F, \\
	& N_c = N'_{c} K',  \\
	& N_s = N'_{s} K,
	\end{align}
\end{subequations}
where $I_{cl} \in \mathbb{Z}^+$, $K' \in \mathbb{Z}^+$ , $N'_{c} \in \mathbb{Z}^+$, $N'_{s} \in \mathbb{Z}^+$, and $N'_{c} > 1$, $N'_{s} > 1$.
That is $M$ BSs form $I_{cl}$ frequency reuse clusters (FRCs), each with $F$ BSs.
The BSs in each FRC utilize the $K$ subcarriers orthogonally, with $K'$ subcarriers being available for each BS.
BS-CU links belonging to the same BS are scheduled in a round-robin manner in the allocated $K'$ subcarriers
via time division multiple access (TDMA). 
For spectrum sharing,
the SU-satellite links are also round-robin scheduled in the $K$ subcarriers via TDMA, but in an opportunistic way.
To depress inter-link interference suffered by BS-CU links,
only one SU is allowed to be served in each subcarrier at each specific moment.
For convenience of presentation,
set $\mathbb{K} = \{1,...,K\}$, $\mathbb{F} = \{ 1,...,F \}$, $\mathbb{I}_{cl} = \{1,...,I_{cl}\}$, 
$\mathbb{N}_c = \{1,...,N_c\}$, and $\mathbb{N}_s = \{1,...,N_s\}$.
The $r$th BS in the $i$th FRC is denoted as $(i,r)$, $i \in \mathbb{I}_{cl}$, $r \in \mathbb{F}$,
and the CU $v$ served by BS $(i,r)$ is denoted as $(i,r,v)$, $i \in \mathbb{I}_{cl}$, $r \in \mathbb{F}$, $v \in \mathbb{N}_c$.

The satellites, BSs, and SUs are all assumed to be equipped with antenna arrays containing multiple antenna elements, 
while each CU is equipped with a single antenna.
Due to down tilt of the BS antenna arrays, interference from BS-CU links to SU-satellite links 
is ignored~\cite{10678835, WangChinacom2018}. 
Correspondingly, the received signals in SU-satellite links and BS-CU links can be respectively written as 
\begin{subequations}\label{eq1}
	\begin{align}
	&y_{u}^{(sat,su)} = \hat{R}_{u}^{(sat)} h_{u}^{(sat,su)} T_{u}^{(su)} x_{u}^{(su)} + n^{(sat)}, \\
	&y_{i,r,v}^{(cu,bs)} = h_{i,r,v}^{(cu,bs)} T_{i,r,v}^{(bs)} x_{i,r,v}^{(bs)} \nonumber \\
	& \,\,\,\,\,\,\,\,\,\,\,\,\,\,\,\,\,\,\,\,\,\,\,\, + {h}_{i,r,v,\hat{u}_{i,r,v}}^{(cu,su)} \hat{T}_{i,r,v,\hat{u}_{i,r,v}}^{(cu,su)} x_{\hat{u}_{i,r,v}}^{(su)} + n^{(cu)},
	\end{align}
\end{subequations}
where $u \in \mathbb{N}_s$, $i \in \mathbb{I}_{cl}$, $r \in \mathbb{F}$, $v \in \mathbb{N}_c$, and $\hat{u}_{i,r,v} \in \mathbb{N}_s$.
$y_{u}^{(sat,su)}$ denotes received signal of the SU-satellite link for SU $u$, 
and $y_{i,r,v}^{(cu,bs)}$ is that of the BS-CU link for CU $(i,r,v)$.
$|\hat{R}_{u}^{(sat)}|^2$ is the receive antenna gain of the satellite serving SU $u$,
$|T_{u}^{(su)}|^2$ represents the transmit antenna gain of SU $u$,
and $|T_{i,r,v}^{(bs)}|^2$ is that of BS $(i,r)$ to CU $(i,r,v)$.
$h_{u}^{(sat,su)}$ is channel of the SU-satellite link for SU $u$, and $h_{i,r,v}^{(cu,bs)}$ is that of the BS-CU link for CU $(i,r,v)$.
$x_{u}^{(su)}$ denotes the transmit signal of SU $u$, while $x_{i,r,v}^{(bs)}$ represents that of BS $(i,r)$ to CU $(i,r,v)$.
$x_{\hat{u}_{i,r,v}}^{(su)}$ denotes transmit signal of the SU $\hat{u}_{i,r,v}$ that 
shares the same subcarrier with CU $(i,r,v)$ simultaneously, 
$|\hat{T}_{i,r,v,\hat{u}_{i,r,v}}^{(cu,su)}|^2$ is the transmit antenna gain of the SU towards $(i,r,v)$,
and ${h}_{k,i,v,\hat{u}_{i,r,v}}^{(cu,su)}$ is the channel from the SU to CU $(i,r,v)$.
$n^{(sat)}$ and $n^{(cu)}$
are white Gaussian noise at the satellites and CUs, respectively.\footnote{Interference between BS-CU links reusing the same subcarriers is not listed separately and could be included in the Gaussian noise if necessary.}
Due to the diversity of propagation environment for the hybrid network,
different channel models could be adopted for the links.
For generality, no special ones are assumed in the system model.

To achieve higher efficiency in spectrum sharing,
scheduling of the BS-CU links is cooperatively optimized with joint scheduling and power control for the SU-satellite links.
To reduce complexity and signaling overhead, 
it is assumed that time synchronization between the satellite and terrestrial components 
is only achieved at a coarse time scale $\Delta \tau$.
It could be much larger than the duration of each ms-level time slot in TDMA for the CUs and SUs.
For flexibility, TDMA in each subcarrier can be implemented independently for the CUs and SUs
within each time interval of $\Delta \tau$.
In other words, the duration of each time slot for the CUs and that for the SUs are set independently.
The scheduling order of the CUs and that of the SUs are independent, too.
Correspondingly, spectrum sharing optimization for BS-CU and SU-satellite links is also implemented at the time scale $\Delta \tau$, and
only statistical CSI of the links across $\Delta \tau$ is utilized.
To achieve appropriate trade-off between complexity and performance in different network scenarios,
$\Delta \tau$ is allowed to be adjusted adaptively according to practical requirements.

\subsection{Problem Formulation}
As serving time of the hybrid network could be divided into time cycles of $\Delta \tau$ for spectrum sharing optimization,
we set
\begin{equation}\label{eq_2}
T = \Delta \tau,
\end{equation}
for convenience of analysis.
Suppose the channel fading statistics of each link is the same in all $K$ subcarriers, and
let $\mathbb{H}_{csi}$ denote statistical CSI of the links within $T$, as
\begin{equation}\label{eq4_0}
\begin{split}
\mathbb{H}_{csi} = &  \{ \eta_{i,r,v}^{(cu,bs)}, \eta_{u}^{(sat,su)}, \eta_{i,r,v,u}^{(cu,su)}   \\
& \,\,\,\,\,\,\,\,\,\,\,  \,|\, i \in \mathbb{I}_c, r \in \mathbb{F}, v \in \mathbb{N}_c, u \in \mathbb{N}_s   \}.
\end{split}
\end{equation} 
$\eta_{i,r,v}^{(cu,bs)}$, $\eta_{u}^{(sat,su)}$, and $\eta_{i,r,v,u}^{(cu,su)}$
denote statistical CSI of the BS-CU, SU-satellite, and SU-CU interference links, respectively.
Based on~(\ref{eq1}) and~(\ref{eq4_0}), the average transmission rate for the CUs and SUs can be expressed as
\begin{subequations}\label{eq5_0}
\begin{align}
\allowdisplaybreaks[4]
&  C^{(cu)}_{i,r,v} = \mathbf{E}_{\mathbb{H}_{csi}} [B \log_2  ( 1+  \frac{|h_{i,r,v}^{(cu,bs)}|^2 |T_{i,r,v}^{(bs)}|^2 P_{bs}}{ \mathfrak{I}_{i,r,v} +{\sigma^{(cu)}}^2 }   ) ],  \\
& C^{(su)}_u = \mathbf{E}_{ \mathbb{H}_{csi}} [ B \log_2  ( 1+ \frac{|\hat{R}_{u}^{(sat)}|^2 |h_{u}^{(sat,su)}|^2 |T_{u}^{(su)}|^2 p_{u}^{(su)} }{{\sigma^{(sat)}}^2} )  ].
\end{align}
\end{subequations}
$\mathbf{E}_{ \mathbb{H}_{csi}} \left[ \cdot \right]$ is the expectation operator with respect to 
channel fading of the links under $\mathbb{H}_{csi}$, i.e., when $\mathbb{H}_{csi}$ is given, and
\begin{equation}\label{eq6_0}
\mathfrak{I}_{i,r,v} = \mathbf{E}_{\mathbb{H}_{csi}} [ |{h}_{i,r,v,\hat{u}_{i,r,v}}^{(cu,su)}|^2 |\hat{T}_{i,r,v,\hat{u}_{i,r,v}}^{(cu,su)}|^2 p_{\hat{u}_{i,r,v}}^{(su)} ], 
\end{equation}
is inter-link interference suffered by CU $(i,r,v)$.
$P_{bs}$ denotes transmit power of the BSs to each CU, and $p_{u}^{(su)}$ is that of SU $u$.
${\sigma^{(cu)}}^2$ and ${\sigma^{(sat)}}^2$ are the power of $n^{(cu)}$ and $n^{(sat)}$, respectively.
Accordingly, the average sum rate of the hybrid network can be written as
\begin{equation}\label{eq10_0}
C_{sum} =  \frac{1}{N'_c}  \sum_{i=1}^{I_{cl}} \sum_{r=1}^{F} \sum_{v=1}^{N_c} C^{(cu)}_{i,r,v} \\
+ \frac{1}{N'_s} \sum_{u=1}^{N_s} C^{(su)}_{u}.
\end{equation}

Let 
\begin{equation}\label{eq13_0}
\mathbb{K}'_{r} = \{(r-1)K'+1,...,rK'\}, r \in \mathbb{F},
\end{equation}
denote $F$ groups of subcarriers allocated to the $F$ BSs in each FRC. 
Further, 
let $\delta^{(cu)}_{i,r,v,k_r} \in \{0,1\}$, $i \in \mathbb{I}_{cl}$, $r \in \mathbb{F}$, $v \in \mathbb{N}_c$, $k_r \in \mathbb{K}'_r$,
represent scheduling indicators for the CUs,
and $\delta_{u,k}^{(su)} \in \{0,1\}$, $u \in \mathbb{N}_s$, $k \in \mathbb{K}$, denote those for the SUs.
If the BS-CU link for CU $(i,r,v)$ is scheduled in subcarrier $k_r$, then $\delta^{(cu)}_{i,r,v,k_r} = 1$, 
and otherwise $\delta^{(cu)}_{i,r,v,k_r} = 0$.
Similarly, if the SU-satellite link for SU $u$ is scheduled in subcarrier $k$, then $\delta^{(su)}_{u,k}=1$,  
and otherwise $\delta^{(su)}_{u,k}=0$. Thus,
\begin{subequations}\label{eq4}
	\begin{align}
\allowdisplaybreaks[4]
	& \sum_{k_r \in \mathbb{K}'_{r} } \delta^{(cu)}_{i,r,v,k_r}  = 1, i \in \mathbb{I}_{cl}, r \in \mathbb{F}, v \in \mathbb{N}_c, \\
	& \sum_{v=1}^{N_c} \delta^{(cu)}_{i,r,v,k_r} = N'_c, i \in \mathbb{I}_{cl}, r \in \mathbb{F},k_r \in \mathbb{K}'_r, \\
	&\sum_{k=1}^{K} \delta^{(su)}_{u,k} = 1, u \in \mathbb{N}_s, \\
	&\sum_{u=1}^{N_s} \delta^{(su)}_{u,k} = N'_s, k \in \mathbb{K}.
	\end{align}
\end{subequations}
As TDMA in each subcarrier is implemented independently for the CUs and SUs within T,
the SU $\hat{u}_{i,r,v}$ that interferes the BS-CU link for CU $(i,r,v)$ can't be known exactly.
Instead, we only get that 
\begin{equation}\label{eq14_0}
\hat{u}_{i,r,v} \in \{ u \,|\, \delta^{(cu)}_{i,r,v,k_r} = 1, \delta^{(su)}_{u,k_r} = 1, k_r \in \mathbb{K}'_{r}, u \in \mathbb{N}_s  \}.
\end{equation}
Accordingly, the worst-case model is adopted
to evaluate inter-link interference at CU $(i,r,v)$, i.e., $\mathfrak{I}_{i,r,v}$ in~(\ref{eq6_0})~\cite{10678835}.
When CU $(i,r,v)$ is served in subcarrier $k_r$, $\mathfrak{I}_{i,r,v}$ is replaced by its upper bound
\begin{equation}\label{eq15_0}
\mathfrak{\bar{I}}_{i,r,v,k_r} = \max_{ u \in \mathbb{N}_s } \left[\delta^{(su)}_{u,k_r} \mathfrak{\hat{I}}_{i,r,v,u}^{(cu,su)}(p_{u}^{(su)}) \right],
\end{equation}
where
\begin{equation}\label{eq15_1}
\mathfrak{\hat{I}}_{i,r,v,u}^{(cu,su)}(p_{u}^{(su)}) = \mathbf{E}_{ \mathbb{H}_{csi}} [ |h_{i,r,v,u}^{(cu,su)}|^2 ] |\hat{T}_{i,r,v,u}^{(cu,su)}|^2  p_{u}^{(su)}.
\end{equation}
Correspondingly, for each $k_r \in \mathbb{K}'_{r}$,
a lower bound for the average transmission rate of CU $(i,r,v)$, i.e., $C^{(cu)}_{i,r,v}$ given in~(\ref{eq5_0}a),
could be written as
\begin{equation}\label{eq17_0}
\underline{C}^{(cu)}_{i,r,v,k_r} = \mathbf{E}_{\mathbb{H}_{csi}} [B \log_2  ( 1+  \frac{|h_{i,r,v}^{(cu,bs)}|^2 |T_{i,r,v}^{(bs)}|^2 P_{bs}}{ \mathfrak{\bar{I}}_{i,r,v,k_r} +{\sigma^{(cu)}}^2 }   ) ].
\end{equation}
Based on~(\ref{eq15_0}) and~(\ref{eq17_0}), a lower bound for $C_{sum}$ given in~(\ref{eq10_0}) can be obtained as
\begin{equation}\label{eq16_0}
\underline{C}_{sum} = \sum_{i=1}^{I_{cl}} \sum_{r=1}^{F} \sum_{v=1}^{N_c} \sum_{k_r \in \mathbb{K}'_{r}} \delta^{(cu)}_{i,r,v,k_r}  \frac{\underline{C}^{(cu)}_{i,r,v,k_r}}{N'_c} +  \sum_{u=1}^{N_s}  \frac{C^{(su)}_{u}}{N'_s}.
\end{equation} 

Suppose $J'_{u}$ of the $J$ satellites are available for SU $u$, $u \in \mathbb{N}_s$, across $T$.
Let $\zeta_{u,j}$, $j = 1,...,J'_u$, denote satellite selection indicators for SU $u$.
If the $j$th available satellite is selected to serve SU $u$, then $\zeta_{u,j} = 1$, and otherwise $\zeta_{u,j} = 0$.
Accordingly, 
\begin{equation}\label{eq7}
\sum_{j=1}^{J'_{u}} \zeta_{u,j} = 1, \,\, u \in \mathbb{N}_s.
\end{equation}
Further, $\hat{R}_{u}^{(sat)}$ and $\hat{T}_{i,r,v,u}^{(cu,su)}$ in~(\ref{eq1}a) and~(\ref{eq1}b) can be respectively expressed as
\begin{subequations}\label{eq8}
\begin{align}
& \hat{R}_{u}^{(sat)} = \sum_{j=1}^{J'_{u}} \zeta_{u,j} R_{u,j}^{(sat)}, \\
& \hat{T}_{i,r,v,u}^{(cu,su)} = \sum_{j=1}^{J'_{u}} \zeta_{u,j} T_{i,r,v,u,j}^{(cu,su)}.
\end{align}
\end{subequations}
$|R_{u,j}^{(sat)}|^2$ is the receive antenna gain of the $j$th available satellite towards SU $u$,
and $|T_{i,r,v,u,j}^{(cu,su)}|^2$ is the off-boresight gain of the transmit antenna array of SU $u$ towards CU $(i,r,v)$ 
when it is served by the $j$th available satellite.
It can be seen from~(\ref{eq5_0}b), (\ref{eq15_0})--(\ref{eq17_0}), and~(\ref{eq8}) that,
with satellite selection, the diversity in the strength of SU-satellite links and that in interference they may cause to CUs
are exploited for opportunistic spectrum sharing.

To guarantee QoS for the CUs and SUs, inter-link interference suffered by all CUs is controlled under
a tolerable threshold $\gamma_{th}$ across $T$, 
and a minimum average transmission rate $\mathcal{C}_u^{QoS}$ is ensured for each SU $u$.
With the target of maximizing $\underline{C}_{sum}$, the spectrum sharing problem for the hybrid network can be formulated as
\begin{subequations}\label{eq18_0}
	\begin{align}
\allowdisplaybreaks[4]
	&\!\!\!\!\!\!\!\!\!\!\!\!\!  \mathop {\max }\limits_{\left\{ \delta^{(cu)}_{i,r,v,k_r}, \delta^{(su)}_{u,k}, \zeta_{u,j}, p_{u}^{(su)} \right\} } \underline{C}_{sum}  \\
	{s.t.} \;\; &(\ref{eq5_0}b), (\ref{eq4}),  (\ref{eq15_0}), (\ref{eq17_0}), (\ref{eq16_0}), (\ref{eq7}), (\ref{eq8}),   \\
	\;\;\;\;\;\; & \sum_{k_r \in \mathbb{K}'_{r}} \delta^{(cu)}_{i,r,v,k_r} \mathfrak{\bar{I}}_{i,r,v,k_r} \leq \gamma_{th}, \, \forall i,r,v, \\
	\;\;\;\;\;\; & C^{(su)}_{u} \geq \mathcal{C}_u^{QoS}, \, \forall u, \\
	\;\;\;\;\;\; &\delta^{(cu)}_{i,r,v,k_r} \in \{0,1\}, \, \forall i,r,v,k_r, \\
	\;\;\;\;\;\; &\delta^{(su)}_{u,k} \in \{0,1\}, \, \zeta_{u,j} \in \{0,1\}, \,  \forall u,k,j, \\
	\;\;\;\;\;\; &0 \leq p_{u}^{(su)} \leq P_{su}, \, \forall u,
	\end{align}
\end{subequations}
where $P_{su}$ is the maximum transmit power of the SUs.

Accurate depiction of channel fading characteristics of the BS-CU, SU-satellite, and SU-CU interference links 
with $\mathbb{H}_{csi}$ presented in~(\ref{eq4_0}) is critical for spectrum sharing based on~(\ref{eq18_0}).
Due to  diversity of the propagation environment in the hybrid network and
random movement of the CUs and SUs, channels of the links may experience various fading processes across $T=\Delta \tau$.
Accordingly, different channel models may be adopted to depict channel fading of the links~\cite{r_TC_1999, r_TVT_2016}. 
Furthermore, 
parameters contained in $\mathbb{H}_{csi}$ should adapt to the duration of $\Delta \tau$,
and they can be acquired with the aid of pilot signals and/or radio maps~\cite{CXWang_1,WangChinacom2018, r_TVT_2016}.
Basically, channels of the links can be seen as 
being composed of large-scale and small-scale fading~\cite{CXWang_1, r_TC_1999, r_TVT_2016}.
The former mainly refers to relatively-slowly-varying path loss and shadowing, 
and the latter represents fast variation of the channels usually caused by multipath propagation~\cite{CXWang_1, r_TC_1999, r_TVT_2016}.
For small-scale fading, only statistical parameters of its distribution, 
such as the Rician K-factor, are known in $\mathbb{H}_{csi}$.
Comparatively, for large-scale fading, either exact values indicating the concrete fading state or statistical parameters 
of its distribution, e.g., means and variances of log-normal-distributed shadowing, 
could be available, depending on the specific duration of $\Delta \tau$.
As a result, it is usually difficult to find unified closed-form, or even relatively-concise, expressions
for $C^{(su)}_{u}$ and $\underline{C}^{(cu)}_{i,r,v,k_r}$ in~(\ref{eq5_0}b) and~(\ref{eq17_0}),
and further for $\underline{C}_{sum}$ in~(\ref{eq16_0}), except for some special cases~\cite{WangChinacom2018, r_JSAC_2021}.

\section{Time-Scale-Adaptable Spectrum Sharing with Link Feature Sketching}

\subsection{Analysis of the Spectrum Sharing Problem}
With link scheduling for both the CUs and SUs and power control for the SUs being complicated coupled together,
(\ref{eq18_0}) is an MIP problem that is challenging to solve~\cite{ref_MIP_problem}.
In the following, we try to uncover some of its solution-friendly characteristics
guided by the divide-and-conquer principle.

\subsubsection{Power Control for the SUs}
For a group of given scheduling indicators for the CUs and SUs, i.e., $\delta^{(cu)}_{i,r,v,k_r}$, 
$\delta^{(su)}_{u,k}$, and $\zeta_{u,j}$, $\forall i, r, v, k_r, u, k$,
set
\begin{subequations}\label{eq19_1}
\begin{align}
&\mathbb{V}_{k_r} = \{ (i,r,v) \,|\, \delta^{(cu)}_{i,r,v,k_r} = 1, i \in \mathbb{I}_{cl}, v \in \mathbb{N}_c  \}, \\
&\mathbb{U}_{k_r} = \{ u \,|\, \delta^{(su)}_{u,k_r} = 1, u \in \mathbb{N}_s \}, 
\end{align}
\end{subequations}
to represent the sets of CUs and SUs scheduled in each subcarrier $k_r$, $k_r \in \mathbb{K}'_r$, $r \in \mathbb{F}$, respectively.
With $\mathbb{V}_{k_r}$, the constraints in~(\ref{eq18_0}c) can be transformed into 
\begin{equation}\label{eq21_1}
\mathfrak{\bar{I}}_{i,r,v,k_r} \leq \gamma_{th}, \,\, (i,r,v) \in \mathbb{V}_{k_r}, r \in \mathbb{F}, k_r \in \mathbb{K}'_r.
\end{equation}
Further, based on~(\ref{eq15_0}) and~(\ref{eq19_1}b), (\ref{eq21_1}) can be rewritten as
\begin{equation}\label{eq22_1}
\begin{aligned}
&\mathfrak{\hat{I}}_{i,r,v,u}^{(cu,su)} (p_{u}^{(su)}) \leq \gamma_{th}, \\
& \,\,\,\,\,\,\,\,\,\,\,\,\,\,\,\,\,\,\,\,\,\,\,\,\,\,\,\,\,\,  (i,r,v) \in \mathbb{V}_{k_r}, u \in \mathbb{U}_{k_r}, k_r \in \mathbb{K}'_r, r \in \mathbb{F},
\end{aligned}
\end{equation}
where $\mathfrak{\hat{I}}_{i,r,v,u}^{(cu,su)} (p_{u}^{(su)})$ is given by~(\ref{eq15_1}).
Correspondingly, power control for the SUs based on~(\ref{eq18_0}) can be equivalently decomposed into $K$ subproblems, 
one for each subcarrier $k_r$, as
\begin{subequations}\label{eq23_1}
	\begin{align}
\allowdisplaybreaks[4]
	&\!\!\!\!\!\!\!\!\!\!\!\!\!  \mathop {\max}\limits_{ \{p_{u}^{(su)},\, t_{i,r,v} \} }  \sum_{(i,r,v) \in \mathbb{V}_{k_r}} \frac{\underline{C}'^{(cu)}_{i,r,v} (t_{i,r,v})}{N'_c}   +   \sum_{ u \in \mathbb{U}_{k_r} } \frac{C^{(su)}_{u}}{N'_s}  \\
	{s.t.} \;\; &\mathfrak{\hat{I}}_{i,r,v,u}^{(cu,su)} (p_{u}^{(su)}) \leq t_{i,r,v}, \,  (i,r,v) \in \mathbb{V}_{k_r}, u \in \mathbb{U}_{k_r},  \\
	\;\;\;\;\;\; & 0 \leq t_{i,r,v} \leq \gamma_{th}, \, (i,r,v) \in \mathbb{V}_{k_r}, \\
	\;\;\;\;\;\; & C^{(su)}_{u} \geq \mathcal{C}_u^{QoS}, \, u \in \mathbb{U}_{k_r}, \\
	\;\;\;\;\;\; & 0 \leq p_{u}^{(su)} \leq P_{su}, \, u \in \mathbb{U}_{k_r},
	\end{align}
\end{subequations}
where $C^{(su)}_{u}$ is given by~(\ref{eq5_0}b) and~(\ref{eq8}a), and
\begin{equation}\label{eq27_1}
\underline{C}'^{(cu)}_{i,r,v} (t_{i,r,v}) = C^{(cu)}_{i,r,v} |_{ \mathfrak{I}_{i,r,v} = t_{i,r,v} },
\end{equation}
with $C^{(cu)}_{i,r,v}$ being given by~(\ref{eq5_0}a), (\ref{eq6_0}) and~(\ref{eq8}b).
Note that compared to~(\ref{eq18_0}),
a group of auxiliary parameters $t_{i,r,v}$, $(i,r,v) \in \mathbb{V}_{k_r}$, are introduced in~(\ref{eq23_1}). 
With the aid of $t_{i,r,v}$, the worst-case interference in~(\ref{eq15_0}) is equivalently replaced
by $\mathfrak{\hat{I}}_{i,r,v,u}^{(cu,su)} (p_{u}^{(su)})$ in~(\ref{eq15_1}),
as indicated by (\ref{eq23_1}b) and (\ref{eq23_1}c).

Due to the non-concavity of $\underline{C}'^{(cu)}_{i,r,v} (t_{i,r,v})$ in~(\ref{eq23_1}a), as shown by~(\ref{eq5_0}a) and~(\ref{eq27_1}),
(\ref{eq23_1}) is a non-convex problem.
However, it is noted that $\underline{C}'^{(cu)}_{i,r,v} (t_{i,r,v})$ can be rewritten as
\begin{equation}\label{eq27_2}
\underline{C}'^{(cu)}_{i,r,v} (t_{i,r,v}) = B [\mathfrak{C}_{i,r,v}^{(1)}(t_{i,r,v}) - \mathfrak{C}_{i,r,v}^{(2)}(t_{i,r,v}) ],
\end{equation}
where 
\begin{subequations}\label{eq27_3}
\begin{align}
\allowdisplaybreaks[4]
& \mathfrak{C}_{i,r,v}^{(1)}(t_{i,r,v}) = \mathbf{E}_{\mathbb{H}_{csi}} [\log_2  ( |h_{i,r,v}^{(cu,bs)}|^2 |T_{i,r,v}^{(bs)}|^2 P_{bs} \nonumber  \\
&     \,\,\,\,\,\,\,\,\,\,\,\,\,\,\,\,\,\,\,\,\,\,\,\,\,\,\,\,\,\,\,\,\,\,\,\,\,\,\,\,\,\,\,\,\,\,\,\,\,\,\,\,\,\,\,\,\,\,\,\,\,\,\,\,\,\,\,\,\,\,\,\,\,\,\,    + t_{i,r,v} +{\sigma^{(cu)}}^2  ) ],  \\
& \mathfrak{C}_{i,r,v}^{(2)}(t_{i,r,v}) = \log_2  (  t_{i,r,v} +{\sigma^{(cu)}}^2 ).
\end{align}
\end{subequations}
It shows that~(\ref{eq23_1}a) is actually a combination of a group of concave functions with respect to $p_{u}^{(su)}$ and $t_{i,r,v}$, 
i.e., $\mathfrak{C}_{i,r,v}^{(1)}(t_{i,r,v})$, $\mathfrak{C}_{i,r,v}^{(2)}(t_{i,r,v})$, and $C^{(su)}_{u}$,
with both positive and negative coefficients.  
Thus, the power control problem in~(\ref{eq23_1}) can be solved based on successive-approximation-aided 
convex optimization~\cite{ref_SuccessiveApproximation}.
Specifically, by successively approximating  each $\mathfrak{C}_{i,r,v}^{(2)}(t_{i,r,v})$ in~(\ref{eq27_3}b)
with its first-order Taylor expansion with respect to $t_{i,r,v}$, 
(\ref{eq23_1}) can be transformed into a series of convex subproblems~\cite{ref_SuccessiveApproximation, ref_cvx_Boyd}.
To deal with the expectation operator $\mathbf{E}_{ \mathbb{H}_{csi}} \left[ \cdot \right]$ in a general sense,
the Monte Carlo method could be utilized, i.e., 
\begin{equation}\label{eq27_4}
	\mathbf{E}_{ \mathbb{H}_{csi}} \left[ \mathfrak{g}(\mathfrak{h}) \right] \approx \frac{1}{Q} \sum_{q=1}^{Q} \mathfrak{g}(\mathfrak{\tilde{h}}_q).
\end{equation}
where 
\begin{equation}\label{eq27_5}
\mathfrak{h} = \{ h_{u}^{(sat,su)}, h_{i,r,v}^{(cu,bs)}, h_{i,r,v,u}^{(cu,su)}  \,|\, \forall u,i,r,v  \},
\end{equation}
denotes channels for all the links, 
$\mathfrak{g}(\mathfrak{h})$ could be $C^{(cu)}_{i,r,v}$, $C^{(su)}_u$ or related expressions that are functions of $\mathfrak{h}$,
and $\mathfrak{\tilde{h}}_q$, $q=1,...,Q$, represent $Q$ random samples for $\mathfrak{h}$ with given $\mathbb{H}_{csi}$.

\subsubsection{Link Scheduling for the CUs and SUs}
When link scheduling for the CUs and SUs is implemented, the impact of power control for the SUs should be considered,
as they are closely coupled in~(\ref{eq18_0}).
Unfortunately, 
link scheduling for~(\ref{eq18_0}) is an intractable NP-hard problem even when all $p_{u}^{(su)}$ are fixed~\cite{ref_NP}.
It prompts us to take a different tack rather than implementing link scheduling by solving~(\ref{eq18_0}) directly. 
Due to coarse satellite-terrestrial time synchronization,
the worst-case model is adopted for interference coordination between BS-CU links and SU-satellite links scheduled in the same subcarrier,
as shown in~(\ref{eq15_0}).
Thus, link scheduling in~(\ref{eq18_0}) is equivalent to cooperatively grouping the BS-CU and SU-satellite links
into $K$ clusters, respectively, one for each subcarrier.
Inspired by our previous work in~\cite{10678835}, we can implement link scheduling for the CUs and SUs
based on link-feature-sketching-aided link clustering.
By delicately-designed feature vectors, BS-CU or SU-satellite links
could be clustered efficiently according to their similarity in interference coordination.
In this way, diversity of the links brought by the spatial distribution of the CUs and SUs
can be reserved in link clusters and fully utilized for opportunistic spectrum sharing.

Specially, the impact of power control for the SUs should be considered in feature sketching for the links.
Without confusion, let $\mathbb{V}_{k}$ and $\mathbb{U}_{k}$, $k=1,...,K$, denote $K$ BS-CU and SU-satellite link clusters, respectively.
Further, without loss of generality, 
BS-CU and SU-satellite links in $\mathbb{V}_{k}$ and $\mathbb{U}_{k}$ are supposed to be scheduled in subcarrier $k$, 
as indicated by~(\ref{eq19_1}).
Let $\underline{p}_{u}^{QoS}$ denote the minimum transmit power of SU $u$ for QoS guarantee, i.e.,
\begin{equation}\label{eq28}
C'^{(su)}_{u}( \underline{p}_{u}^{QoS} ) = \mathcal{C}_u^{QoS},
\end{equation}
where 
\begin{equation}\label{eq29}
C'^{(su)}_{u}( \underline{p}_{u}^{QoS} ) = C^{(su)}_u |_{p_{u}^{(su)} = \underline{p}_{u}^{QoS}},
\end{equation}
and $C^{(su)}_u$ is given by~(\ref{eq5_0}b).
Define
\begin{subequations}\label{eq30}
	\begin{align}	
	& \Delta C^{(su)}_{u} (p_{u}^{(su)}) = C'^{(su)}_{u}( p_{u}^{(su)} ) - C'^{(su)}_{u}( \underline{p}_{u}^{QoS} ), \\
	& \Delta C^{(cu)}_{i,r,v} (t_{i,r,v}) = \underline{C}'^{(cu)}_{i,r,v} (t_{i,r,v}) - \underline{C}'^{(cu)}_{i,r,v} (\gamma_{th}).
	\end{align}
\end{subequations}
Set
\begin{equation}\label{eq30_1}
\underline{C}_{sum}^{min} = \sum_{i=1}^{I_{cl}} \sum_{r=1}^{F} \sum_{v=1}^{N_c} \frac{\underline{C}'^{(cu)}_{i,r,v} (\gamma_{th})}{N'_c} + \sum_{u =1}^{N_s} \frac{C'^{(su)}_{u}( \underline{p}_{u}^{QoS} )}{N'_s},
\end{equation}
and for $k=1,...,K$, set
\begin{subequations}\label{eq31}
	\begin{align}
	&\underline{\gamma}_{i,r,v}^{(k,min)} = \max_{u \in \mathbb{U}_{k}} \mathfrak{\hat{I}}_{i,r,v,u}^{(cu,su)} (\underline{p}_{u}^{QoS}), \,\, (i,r,v) \in \mathbb{V}_{k},\\
	&\bar{p}_{u}^{(k,max)} = \min \{ \min_{(i,r,v) \in \mathbb{V}_{k}} \bar{p}_{u}^{(i,r,v)}, \, P_{su} \}, \,\, u \in \mathbb{U}_{k},
	\end{align}
\end{subequations}
where $\bar{p}_{u}^{(i,r,v)}$ is determined by
\begin{equation}\label{eq32_0}
\mathfrak{\hat{I}}_{i,r,v,u}^{(cu,su)} (\bar{p}_{u}^{(i,r,v)}) = \gamma_{th}.
\end{equation}
Suppose $\zeta_{u,j}$, $\delta^{(cu)}_{i,r,v,k_r}$, and $\delta^{(su)}_{u,k}$ are 
a group of feasible scheduling indicators for~(\ref{eq18_0}),
and $\underline{C}_{sum}^{*}$ denotes the maximum $\underline{C}_{sum}$ that could be achieved by power control.
Then it can be inferred that 
\begin{subequations}\label{eq33}
	\begin{align}	
		& \underline{C}_{sum}^{*} \geq  \sum_{k=1}^{K} \sum_{ u \in \mathbb{U}_{k} } \frac{\Delta C^{(su)}_{u} (\bar{p}_{u}^{(k,max)})}{N'_s} + \underline{C}_{sum}^{min}, \\
		& \underline{C}_{sum}^{*} \geq \sum_{k=1}^{K} \sum_{(i,r,v) \in \mathbb{V}_{k}} \frac{\Delta C^{(cu)}_{i,r,v} (\underline{\gamma}_{i,r,v}^{(k,min)})}{N'_c} + \underline{C}_{sum}^{min}.
	\end{align}
\end{subequations}

The right side of~(\ref{eq33}a) and that of~(\ref{eq33}b)
actually constitute two lower bounds for the maximum $\underline{C}_{sum}$ 
promised by spectrum sharing optimization in~(\ref{eq18_0}), i.e., joint link scheduling and power control.
In~(\ref{eq33}a) and~(\ref{eq33}b),
$\underline{C}_{sum}^{min}$ is a fixed part that needs to be achieved for QoS of the CUs and SUs, as indicated by~(\ref{eq30_1}).
Contrarily, $\Delta C^{(su)}_{u} (\bar{p}_{u}^{(k,max)})$ and $\Delta C^{(cu)}_{i,r,v} (\underline{\gamma}_{i,r,v}^{(k,min)})$,
given by~(\ref{eq30}) and~(\ref{eq31}),
depend on quality of the BS-CU and SU-satellite links as well as strength of the SU-CU interference links,
and vary with link scheduling.
By setting $p_{u}^{(su)}$, $u \in \mathbb{U}_{k}$, $k \in \mathbb{K}$, at the maximum and minimum feasible points,
$\Delta C^{(su)}_{u} (\bar{p}_{u}^{(k,max)})$ and $\Delta C^{(cu)}_{i,r,v} (\underline{\gamma}_{i,r,v}^{(k,min)})$
are relieved from complicated coupling between link scheduling and power control
while retaining the ability to indicate performance potential of power control.
Thus, they can be utilized in feature sketching of the links for scheduling.

\subsection{Hierarchical Link Clustering with Link Feature Sketching}
With a frequency reuse factor $F$ for the BS-CU links,
the CUs $(i,r,v)$, $i \in \mathbb{I}_{cl}$, $v \in \mathbb{N}_c$,
share a common set of subcarriers in $\mathbb{K}'_r$ given by~(\ref{eq13_0}), for all $r \in \mathbb{F}$.
Thus, the BS-CU links are naturally divided into $F$ coarse groups according to allocated subcarriers.
For efficient spectrum sharing, clustering of the SU-satellite links should match with the $F$ coarse groups of BS-CU links.
Towards this goal, hierarchical clustering of SU-satellite links is explored in the following.

\subsubsection{Link Feature Sketching}
For clustering of the SU-satellite links, feature vectors are firstly designed for them.
Intuitively, to reserve diversity of the links for joint link scheduling and power control,
SU-satellite links that are similar in quality as well as inter-link interference to BU-CU links could be clustered together.
Based on the lower bounds of $\underline{C}_{sum}$ given in~(\ref{eq33}),
the feature vector for the SU-satellite link for SU $u$ is expressed as
\begin{equation}\label{eq35}
\mathcal{F}_{u} = \left[ \mathcal{F}^{sub}_{u,1}, ..., \mathcal{F}^{sub}_{u,F}  \right],
\end{equation}
where each sub-feature-vector $\mathcal{F}^{sub}_{u,r} $ is set as
\begin{equation}\label{eq35_1}
\mathcal{F}^{sub}_{u,r} = \left[ \mathbf{f}_{1,r,1}^{(u)}, ..., \mathbf{f}_{1,r,N_c}^{(u)}, \mathbf{f}_{2,r,1}^{(u)}, ..., \mathbf{f}_{2,r,N_c}^{(u)}, ...,  \mathbf{f}_{I_{cl},r,N_c}^{(u)}  \right],
\end{equation}
and
\begin{equation}\label{eq36}
\mathbf{f}_{i,r,v}^{(u)} = [ \frac{\Delta C^{(su)}_{u} (\bar{p}_{u}^{(i,r,v)})}{N'_s}, \, \frac{\Delta C^{(cu)}_{i,r,v} (\underline{\gamma}_{i,r,v}^{(u)})}{N'_c}].
\end{equation}
$\Delta C^{(su)}_{u} (\cdot)$ and $\Delta C^{(cu)}_{i,r,v} (\cdot)$ are presented in~(\ref{eq30}a) and~(\ref{eq30}b), respectively,
$\bar{p}_{u}^{(i,r,v)}$ is determined by~(\ref{eq32_0}), and 
\begin{equation}\label{eq37}
\underline{\gamma}_{i,r,v}^{(u)} = \mathfrak{\hat{I}}_{i,r,v,u}^{(cu,su)} (\underline{p}_{u}^{QoS}),
\end{equation}
where $\underline{p}_{u}^{QoS}$ is given by~(\ref{eq28}).
With $\mathcal{F}_{u}$, $u \in \mathbb{N}_s$, the similarity between SU-satellite links for any two SUs $u_1$ and $u_2$ can be measured 
based on the distance between $\mathcal{F}_{u_1}$ and $\mathcal{F}_{u_2}$, as
%\begin{equation}\label{eq38}
%\mathfrak{D}_{u_1, u_2} = ||\mathcal{F}_{u_1} - \mathcal{F}_{u_2}||_1,
%\end{equation}
\begin{equation}\label{eq38}
	\mathfrak{D}( \mathcal{F}_{u_1}, \mathcal{F}_{u_2} ) = ||\mathcal{F}_{u_1} - \mathcal{F}_{u_2}||_1,
\end{equation}
where $||\cdot||_1$ represents the $L_1$ norm~\cite{L1_norm}.

It can be observed from~(\ref{eq35})--(\ref{eq37}) that $\mathcal{F}_{u}$ 
is defined following the optimization target of~(\ref{eq18_0}), i.e., $\underline{C}_{sum}$, and the lower bounds in~(\ref{eq33}).
Specially, $\bar{p}_{u}^{(i,r,v)}$ and $\underline{\gamma}_{i,r,v}^{(u)}$ 
for $\Delta C^{(su)}_{u} (\bar{p}_{u}^{(i,r,v)})$ and $\Delta C^{(cu)}_{i,r,v} (\underline{\gamma}_{i,r,v}^{(u)})$
in each $\mathbf{f}_{i,r,v}^{(u)}$ given by~(\ref{eq36})
are derived based on the inter-link interference between SU $u$ and CU $(i,r,v)$. 

\begin{algorithm}[t]
	\caption{Hierarchical clustering of the SU-satellite links}
	\begin{algorithmic}[1]
		\small
		\label{SUClusteringAlgorithm}
		\STATE \textbf{Coarse link clustering:}
		\STATE Solve the standard assignment problem in~(\ref{eq39_2}), and denote its optimal solution as 
		$\{ \theta^*_{u,r,m} \, |\, \forall u, r,m \}$.
		\STATE Group the SU-satellite links into $F$ coarse clusters
		$\mathbb{\hat{S}}_r^{*(su)}$, $r \in \mathbb{F}$, based on~(\ref{eq40_2}a), and get the satellite selection indicators
		$\{\hat{\zeta}_{u,j}^{*} | \forall j \}$ for each SU $u \in \mathbb{N}_s$, as shown in~(\ref{eq40_2}b).    
		\STATE \textbf{Fine link clustering:}
		%		\STATE Set $\mathbb{U}_{k_r} = \phi$, $k_r \in \mathbb{K}'_r$, $r \in \mathbb{F}$.
		\FOR{$r=1,..., F$}
		\STATE Find $\{ \tilde{u}_1, \tilde{u}_2 \} = \arg \max_{u_1, u_2 \in \mathbb{\hat{S}}_r^{*(su)}} \mathfrak{D}(\mathcal{F}^{sub}_{u_1,r}, \mathcal{F}^{sub}_{u_2,r})  $.
		\FOR{$i=3,..., K'$}
		\STATE Set $\mathbb{\hat{S}}'_r = \mathbb{\hat{S}}_r^{*(su)} \backslash \{ \tilde{u}_1, ..., \tilde{u}_{i-1} \}$.
		\STATE Find $\tilde{u}_i = \arg \max_{u_i \in \mathbb{\hat{S}}'_r} \mathfrak{\bar{D}}^{\text{CMPD}}_{u_i, \tilde{u}_1, ..., \tilde{u}_{i-1}}$ based on~(\ref{eq40_4}).
		\ENDFOR
		\STATE Set $\mathcal{F}^{sub}_{\tilde{u}_i,r}$, $i=1,...,K'$, as the initial centers of $K'$ fine SU-satellite link clusters
		$\mathbb{U}_{(r-1)K'+i}$, $i=1,...,K'$, and denote them as $\mathfrak{c}_i = \mathcal{F}^{sub}_{\tilde{u}_i,r}$, $i=1,...,K'$.
		\STATE Set $\mathcal{D}_0^{sum}=0$, $\mathcal{D}_1^{sum} = \sum_{u \in \mathbb{\hat{S}}_r^{*(su)}} ||\mathcal{F}^{sub}_{u,r}||_1$, 
		and $\epsilon = 10^{-2}$.
		\STATE Set $\mathfrak{L}_s = 0$.
		\WHILE{ $|\mathcal{D}_0^{sum}- \mathcal{D}_1^{sum}|/\mathcal{D}_1^{sum} > \epsilon $ }
		\STATE Set $\mathcal{D}_0^{sum} = \mathcal{D}_1^{sum}$ and $\mathcal{D}_1^{sum}=0$.
		\STATE Set $\Phi = \mathbb{\hat{S}}_r^{*(su)}$, $\mathbb{U}_{k_r} = \phi$, $k_r \in \mathbb{K}'_r$, and $\Phi' = \{1,...,K'\}$.
		\WHILE{ $\Phi \neq \phi$ }
		\STATE Find $\{ \tilde{u}, \tilde{i} \} = \arg \min_{u \in \Phi, i \in \Phi' } || \mathcal{F}^{sub}_{u,r} - \mathfrak{c}_i ||_1$.
		\STATE Group SU $\tilde{u}$ into the fine link cluster $\mathbb{U}_{(r-1)K'+\tilde{i}}$, 
		i.e., set $\mathbb{U}_{(r-1)K'+\tilde{i}} := \mathbb{U}_{(r-1)K'+\tilde{i}} \cup \{ \tilde{u} \}$ and 
		$\Phi := \Phi \backslash \{ \tilde{u} \} $.
		\STATE Let $\mathcal{D}_1^{sum} := \mathcal{D}_1^{sum} + || \mathcal{F}_{\tilde{u}} - \mathfrak{c}_{\tilde{i}} ||_1$.
		\IF{$|\mathbb{U}_{(r-1)K'+\tilde{i}}| = N'_s$}
		\STATE Set $\Phi' := \Phi' \backslash \{ \tilde{i} \} $.
		\ENDIF
		\ENDWHILE
		\STATE Update $\mathfrak{c}_i = 1/N'_s \sum_{u \in \mathbb{U}_{(r-1)K'+i}} \mathcal{F}^{sub}_{u,r}$, $i=1,...,K'$.
		\STATE Set $\mathfrak{L}_s := \mathfrak{L}_s + 1$.
		\ENDWHILE
		\ENDFOR
	\end{algorithmic}
\end{algorithm}

\subsubsection{Hierarchical Link Clustering}
Based on the feature vectors defined in~(\ref{eq35})--(\ref{eq36}),
an algorithm is proposed for hierarchical clustering of the SU-satellite links.
Specially, coarse and fine link clustering are implemented successively, 
as illustrated in Algorithm~\ref{SUClusteringAlgorithm}.

Firstly, guiding by $F$ BS-CU link groups, SU-satellite links are firstly grouped into $F$ coarse clusters, 
which are scheduled in the $F$ sets of subcarriers, $\mathbb{K}'_r$, $r \in \mathbb{F}$, respectively.
Let $\mathbb{S}_r^{(su)}$, $r \in \mathbb{F}$, denote $F$ coarse SU-satellite link clusters, as
\begin{equation}\label{eq38_1}
\mathbb{S}_r^{(su)} = \{ u \,|\, u \in \mathbb{N}_s, \sum_{k \in \mathbb{K}'_r} \delta^{(su)}_{u,k} = 1  \} , \, r \in \mathbb{F}.
\end{equation}
Set
\begin{equation}\label{eq39_0}
\Omega = \{ \mathbb{S}_r^{(su)}, \zeta_{u,j} \,|\, \forall r, u, j \},
\end{equation}
and define
\begin{equation}\label{eq39}
\mathcal{U}^{(su)} (\Omega) =  \sum_{r=1}^F \sum_{u \in \mathbb{S}_r^{(su)}} \sum_{i=1}^{I_{cl}} \sum_{v=1}^{N_c} \mathbf{f}_{i,r,v}^{(u)} [\omega_1,1]^T,
\end{equation}
where $\omega_1 = \frac{N_s/F}{I_{cl} N_c}$ is to balance the average transmission rates of the SUs and CUs in $\mathcal{U}^{(su)} (\Omega)$
referring to $\underline{C}_{sum}$ given in~(\ref{eq16_0}).
It can be inferred from~(\ref{eq16_0}), (\ref{eq33}), and~(\ref{eq36}) that 
SU-satellite link clusters resulting in larger $\mathcal{U}^{(su)} (\Omega)$
tend to lead to larger $\underline{C}_{sum}$.
Correspondingly, $F$ coarse SU-satellite link clusters could be obtained by maximizing $\mathcal{U}^{(su)} (\Omega)$, as
\begin{subequations}\label{eq39_1}
	\begin{align}
\allowdisplaybreaks[4]
	&\!\!\!\!\!\!\!\!\!\!\!\!\!  \mathop {\max }\limits_{\Omega } \,\, \mathcal{U}^{(su)} (\Omega)  \\
	{s.t.} \;\; & (\ref{eq5_0}b), (\ref{eq4}\text{c}), (\ref{eq4}\text{d}), (\ref{eq15_1}), (\ref{eq7}), (\ref{eq8}), (\ref{eq18_0}\text{f}), \\
	\;\;\;\;\;\; & (\ref{eq29}), (\ref{eq30}), (\ref{eq32_0}), (\ref{eq36}), (\ref{eq37}), (\ref{eq38_1}), (\ref{eq39_0}).
	\end{align}
\end{subequations}
It can be shown that~(\ref{eq39_1}) could be solved with polynomial complexity through standard assignment optimization~\cite{ref_AP}.
Specifically, suppose $\{ \theta^*_{u,r,m} \, |\, \forall u, r,m \}$ is an optimal solution for the following standard assignment problem as
\begin{subequations}\label{eq39_2}
	\begin{align}
\allowdisplaybreaks[4]
	&\!\!\!\!\!\!\!\!\!\!\!\!\!  \mathop {\max }\limits_{ \{\theta_{u,r,m}\} } \,\, \sum_{u=1}^{N_s} \sum_{r=1}^F \sum_{m=1}^{N'_s K'} \theta_{u,r,m} \bar{\mathfrak{f}}^{(sum)}_{u,r}  \\
	{s.t.} \;\; & \sum_{r=1}^F \sum_{m=1}^{N'_s K'} \theta_{u,r,m} = 1, \, \forall u, \\
	\;\;\;\;\;\; & \sum_{u=1}^{N_s} \theta_{u,r,m} = 1, \, \forall r, m,
	\end{align}
\end{subequations}
with $\bar{\mathfrak{f}}^{(sum)}_{u,r} = \max_{\{ \zeta_{u,j} | \forall j \}} \sum_{i=1}^{I_{cl}} \sum_{v=1}^{N_c} \mathbf{f}_{i,r,v}^{(u)} [\omega_1,1]^T$.
Set
\begin{subequations}\label{eq40_2}
\begin{align}
\allowdisplaybreaks[4]
& \mathbb{\hat{S}}_r^{*(su)} = \{ u \,|\, u \in \mathbb{N}_s, \sum_{m=1}^{N'_s K'} \theta^*_{u,r,m} = 1  \}, \, r \in \mathbb{F}, \\
& \{\hat{\zeta}_{u,j}^{*} | \forall j \} = \arg \max_{\{ \zeta_{u,j} | \forall j \}} \sum_{i=1}^{I_{cl}} \sum_{v=1}^{N_c} \mathbf{f}_{i,\hat{r}_u^*,v}^{(u)} [\omega_1,1]^T, u \in \mathbb{N}_s,
\end{align}
\end{subequations}
where $\hat{r}_u^* \in \mathbb{F}$ satisfies $u \in \mathbb{\hat{S}}_{\hat{r}_u^*}^{*(su)}$. Then $\hat{\Omega}^* = \{ \mathbb{\hat{S}}_r^{*(su)}, \hat{\zeta}_{u,j}^{*} \,|\, \forall r, u, j \}$
constitutes an optimal solution for~(\ref{eq39_1}).
Actually, $\{ \theta^*_{u,r,m} \, |\, \forall u, r,m \}$ can be seen as a specific 
scheduling order of the SU-satellite links in the subcarriers corresponding to the $F$ coarse clusters, 
$\mathbb{\hat{S}}_r^{*(su)}$, $r \in \mathbb{F}$. 

Next, the SU-satellite links within each $\mathbb{\hat{S}}_r^{*(su)}$, $r \in \mathbb{F}$, 
are further clustered into $K'$ fine clusters, 
denoted as $\mathbb{U}_{k_r}$, $k_r \in \mathbb{K}'_r$, respectively, as presented in~(\ref{eq19_1}b).
For fine link clustering, a modified K-means method is adopted with the aid of 
the sub-feature-vectors and the distance defined in~(\ref{eq35_1}) and~(\ref{eq38}), respectively~\cite{10678835,ref_K_means}.
For each $r \in \mathbb{F}$, main steps for fine link clustering are as follows~\cite{10678835}.
\begin{itemize}
\item[a)] Select $K'$ of the sub-feature-vectors, $\mathcal{F}^{sub}_{u,r}$, $u \in \mathbb{\hat{S}}_r^{*(su)}$, 
for the SU-satellite links in $\mathbb{\hat{S}}_r^{*(su)}$
as initial centers for $K'$ fine link clusters, $\mathbb{U}_{k_r}$, $k_r \in \mathbb{K}'_r$.
\item[b)] Calculate distances between each $\mathcal{F}^{sub}_{u,r}$, $u \in \mathbb{\hat{S}}_r^{*(su)}$, 
and the centers of all $\mathbb{U}_{k_r}$, $k_r \in \mathbb{K}'_r$, and cluster the SU-satellite links into the $K'$ fine link clusters
following the \emph{smallest-distance-first} principle~\cite{10678835}, until there are $N'_s$ SU-satellite links in each $\mathbb{U}_{k_r}$. 
\item[c)] Update the center of each $\mathbb{U}_{k_r}$ as average of the sub-feature-vectors for the SU-satellite links in it,
i.e., $1/N'_s \sum_{u \in \mathbb{U}_{k_r}} \mathcal{F}^{sub}_{u,r} $, and repeat b) until convergence.
\end{itemize}
For selection of the initial centers in Step a), 
a compound distance that could reflect the distance from one sub-feature-vector, $\mathcal{F}^{sub}_{u_1,r}$, to multiple others, $\mathcal{F}^{sub}_{u'_1,r}, \mathcal{F}^{sub}_{u'_2,r}, ..., \mathcal{F}^{sub}_{u'_n,r}$, 
is adopted, as
\begin{equation}\label{eq40_4}
\mathfrak{\bar{D}}^{\text{CMPD}}_{u_1, u'_1, ..., u'_n} = \mathfrak{D}_{u_1, u'_1} \cdot \mathfrak{D}_{u_1, u'_2} \cdot \ldots \cdot  \mathfrak{D}_{u_1, u'_n}.
\end{equation}
where $\mathfrak{D}_{u_1, u'_{\bar{n}}} = \mathfrak{D}(\mathcal{F}^{sub}_{u_1,r}, \mathcal{F}^{sub}_{u'_{\bar{n}},r})$,
$\bar{n} = 1,...,n$.
Based on $\mathfrak{\bar{D}}^{\text{CMPD}}_{u_1, u'_1, ..., u'_n}$, 
a greedy method designed following~\cite{10678835} 
is utilized to find $K'$ sub-feature-vectors with distances as large as possible to each other
to act as initial centers for $K'$ fine link clusters, $\mathbb{U}_{k_r}$, $k_r \in \mathbb{K}'_r$.

\begin{algorithm}[t]
	\caption{A time-scale-adaptable spectrum sharing scheme}
	\begin{algorithmic}[1]
		\small
		\label{SpectrumSharingScheme}
		\STATE \textbf{Link scheduling for the SUs:}
		\STATE Implement Algorithm~\ref{SUClusteringAlgorithm} and group the SU-satellite links into $K$ link clusters, $\mathbb{U}_{k}$, $k=1,...,K$.
		\STATE Obtain a group of scheduling and satellite selection indicators 
		i.e., $\hat{\delta}^{*(su)}_{u,k}$, $\hat{\zeta}^*_{u,j}$, $j=1,...,J'_u$, $u \in \mathbb{N}_s$, $k \in \mathbb{K}$,
		for the SUs based on~(\ref{eq19_1}b) and~(\ref{eq40_2}b), respectively.
		\STATE \textbf{Link scheduling for the CUs:}
		\STATE Solve $M=I_{cl}F$ assignment problems as shown in~(\ref{eq41_2}) and get a group of scheduling indicators for the CUs,
		i.e., $\hat{\delta}^{*(cu)}_{i,r,v,k_r}$, $i \in \mathbb{I}_{cl}$, $r \in \mathbb{F}$, $v \in \mathbb{N}_c$, 
		$k_r \in \mathbb{K}'_r$.
		\STATE Form $K$ BS-CU link clusters, $\mathbb{V}_{k}$, $k=1,...,K$,
		based on~(\ref{eq19_1}a) with $\hat{\delta}^{*(cu)}_{i,r,v,k_r}$, $\forall i,r,v,k_r$.
		\STATE \textbf{Power control for the SUs:}
		\STATE Set $\epsilon = 10^{-2}$.
		\FOR{$ k=1,...,K$}
		\STATE Set $\tilde{p}_u^{(su)} = 0$ and $\tilde{p}_u^{*(su)} = \underline{p}_u^{QoS}$ based on~(\ref{eq28}), 
		$u \in \mathbb{U}_{k}$.
		\STATE Set $\tilde{t}_{i,r,v} = \max_{u \in \mathbb{U}_{k}} \mathfrak{\hat{I}}_{i,r,v,u}^{(cu,su)} (\tilde{p}_u^{*(su)})$, 
		       $\forall i,r,v$, based on~(\ref{eq15_1}).
		\STATE Set $\mathfrak{L}_p = 0$.
		\WHILE{ $ \max_{u \in \mathbb{U}_{k}} | \tilde{p}_u^{*(su)} -\tilde{p}_u^{(su)} |/\tilde{p}_u^{*(su)}  > \epsilon$ }
		\STATE Let $\tilde{p}_{u}^{(su)} = \tilde{p}_{u}^{*(su)}$.
		\STATE Approximate each $\underline{C}'^{(cu)}_{i,r,v} (t_{i,r,v})$ in~(\ref{eq23_1}a) with~(\ref{eq53}), 
		and solve the approximated convex problem of~(\ref{eq23_1}). Denote the optimal solution as $\tilde{p}_{u}^{*(su)}$, $\tilde{t}_{i,r,v}$, 
		$\forall u, i, r, v $.
		\STATE Set $\mathfrak{L}_p := \mathfrak{L}_p + 1$.
		\ENDWHILE
		\ENDFOR
		\STATE Output $\hat{\delta}^{*(su)}_{u,k}$, $\hat{\zeta}^*_{u,j}$, $\hat{\delta}^{*(cu)}_{i,r,v,k_r}$, and $\tilde{p}_{u}^{*(su)}$,
		$\forall u,k,j,i,r,v,k_r$.
	\end{algorithmic}
\end{algorithm}

\subsection{A Time-Scale-Adaptable Spectrum Sharing Scheme}
Based on illustrations in Subsection III-A and Subsection III-B,
a time-scale-adaptable spectrum sharing scheme is proposed for the hybrid network based on link feature sketching.
As illustrated in Algorithm~\ref{SpectrumSharingScheme}, the proposed scheme consists of three parts, 
i.e., 1) link scheduling for the SUs via hierarchical clustering of the SU-satellite links, 
2) link scheduling for the CUs, and 3) power control for the SUs.

Firstly, $K$ SU-satellite link clusters, i.e., $\mathbb{U}_{k}$, $k=1,...,K$,
and a group of satellite selection indicators for the SUs, $\hat{\zeta}^*_{u,j}$, $u \in \mathbb{N}_s$, $j=1,...,J'_u$, are obtained
via Algorithm~\ref{SUClusteringAlgorithm}.
Accordingly, a group of link scheduling indicators for the SUs, $\hat{\delta}^{*(su)}_{u,k}$, $u \in \mathbb{N}_s$, $k \in \mathbb{K}$, 
could be derived from $\mathbb{U}_{k}$, $k=1,...,K$, based on~(\ref{eq19_1}b), i.e., $\hat{\delta}^{*(su)}_{u,k} = 1$ 
if $u \in \mathbb{U}_{k}$ and $\hat{\delta}^{*(su)}_{u,k} = 0$ otherwise.

Then link scheduling for the CUs is implemented based on $\hat{\delta}^{*(su)}_{u,k}$ and $\hat{\zeta}^*_{u,j}$, $\forall u,k,j$.
With $K$ fine clusters of the SU-satellite links being respectively scheduled in $K$ subcarriers, 
the BS-CU links corresponding to each BS can be scheduled independently. 
To circumvent the NP-hardness of solving~(\ref{eq18_0}) directly,
$M=I_{cl}F$ assignment problems similar to~(\ref{eq39_2}) are formulated for BS-CU link scheduling
using the lower bounds of $\underline{C}_{sum}$ given in~(\ref{eq33}).
For each pair of $i\in \mathbb{I}_{cl}$ and $r \in \mathbb{F}$, the assignment problem can be written as
\begin{subequations}\label{eq41_2}
	\begin{align}
\allowdisplaybreaks[4]
	&\!\!\!\!\!\!\!\!\!\!\!\!\!  \mathop {\max }\limits_{\left\{ \delta^{(cu)}_{i,r,v,k_r} | \forall v, k_r \right\} } \sum_{v=1}^{N_c}  \sum_{k_r \in \mathbb{K}'_{r}} \delta^{(cu)}_{i,r,v,k_r} \mu_{i,r,v,k_r}^{(cu)}  \\
	{s.t.} \;\; & \sum_{k_r \in \mathbb{K}'_{r} } \delta^{(cu)}_{i,r,v,k_r}  = 1, \, v \in \mathbb{N}_c, \\
	\;\;\;\;\;\; & \sum_{v=1}^{N_c} \delta^{(cu)}_{i,r,v,k_r} = N'_c, \, k_r \in \mathbb{K}'_r, \\
	\;\;\;\;\;\; &\delta^{(cu)}_{i,r,v,k_r} \in \{0,1\}, \, \forall v,k_r,
	\end{align}
\end{subequations}
where
\begin{equation}\label{eq41_4}
	\mu_{i,r,v,k_r}^{(cu)} = \left \{
	\begin{aligned}
		& \omega_2 \sum_{u \in \mathbb{U}_{k_r}} \frac{\Delta C^{(su)}_{u} (\bar{p}_{u}^{(i,r,v)})}{ N'_s} +  \frac{\Delta C^{(cu)}_{i,r,v} (\underline{\gamma}_{i,r,v}^{(k_r,min)})}{ N'_c}, \\
		& \,\,\,\,\,\,\,\,\,\,\,\,\,\,\,\,\,\,\,\,\,\,\,\,\,\,\,\,\,\,\,\,\,\,\,\,\,\,\,\,\,\,\,\,\,\,\,\,\,\,\,\,\,\,\,\,\,\,\,\,\,\,\,\,\,\,\,\,\,\,\,\,\,\, \text{if $\underline{\gamma}_{i,r,v}^{(k_r,min)} < \gamma_{th}$},\\
		& -\mathcal{M}, \,\, \text{otherwise}.
	\end{aligned}
	\right.
\end{equation}
$\underline{\gamma}_{i,r,v}^{(k,min)}$ and $\bar{p}_{u}^{(i,r,v)}$ are given by~(\ref{eq31}a) and~(\ref{eq32_0}), respectively. 
$\mathcal{M}$ is a positive number large enough to 
ensure that QoS for the CUs and SUs in~(\ref{eq18_0}c) and~(\ref{eq18_0}d) is satisfied while maximizing~(\ref{eq41_2}a).
$\omega_2 = 1/(I_{cl}N'_c) $ is to balance average transmission rates of the SUs and CUs 
referring to $\underline{C}_{sum}$ given in~(\ref{eq16_0}).
Each problem in~(\ref{eq41_2}) can be solved with polynomial complexity via standard assignment optimization 
like~(\ref{eq39_2})~\cite{ref_AP}.

Lastly, power control for the SUs is implemented in each subcarrier $k \in \mathbb{K}$ 
based on~(\ref{eq23_1}) with the aid of the Monte Carlo method as shown in~(\ref{eq27_4}) and successive approximation.
Following~(\ref{eq27_2}) and~(\ref{eq27_3}), each $\underline{C}'^{(cu)}_{i,r,v} (t_{i,r,v})$ in~(\ref{eq23_1}a) is successively approximated
in each iteration by
\begin{equation}\label{eq53}
\underline{C}'^{(cu)}_{i,r,v} (t_{i,r,v}) \approx B [\mathfrak{C}_{i,r,v}^{(1)}(t_{i,r,v}) - \mathfrak{\tilde{C}}_{i,r,v}^{(2)}(t_{i,r,v}) ],
\end{equation}
where 
\begin{equation}\label{eq54}
\mathfrak{\tilde{C}}_{i,r,v}^{(2)}(t_{i,r,v}) = \mathfrak{C}_{i,r,v}^{(2)}(\tilde{t}_{i,r,v}) + \log_2 e \frac{t_{i,r,v} -\tilde{t}_{i,r,v} }{\tilde{t}_{i,r,v} +{\sigma^{(cu)}}^2},
\end{equation}
is the first-order Taylor expansion of $\mathfrak{C}_{i,r,v}^{(2)}(t_{i,r,v})$ at $\tilde{t}_{i,r,v}$~\cite{ref_SuccessiveApproximation, ref_cvx_Boyd}.
Using this approximation, (\ref{eq23_1}) 
can be solved via convex optimization algorithms, e.g., the interior-point method~\cite{ref_cvx_Boyd}.
A group of random samples for channels of the links, i.e., $\mathfrak{\tilde{h}}_q$, $q=1,...,Q$, 
are needed to deal with the expectation operator $\mathbf{E}_{ \mathbb{H}_{csi}} \left[ \cdot \right]$ 
in the target of~(\ref{eq23_1}) based on~(\ref{eq27_4}).
For each time interval of $T=\Delta \tau$,
they can be generated just once based on the given set of statistical CSI in $\mathbb{H}_{csi}$,
and utilized across the whole optimization process.
Generally, the parameters in $\mathbb{H}_{csi}$, i.e.,
$\eta_{i,r,v}^{(cu,bs)}$, $\eta_{u}^{(sat,su)}$, and $\eta_{i,r,v,u}^{(cu,su)}$ in~(\ref{eq4_0}),
specify the statistical distribution of channel fading for the links within $T=\Delta \tau$, 
as illustrated at the end of Section II-B. 
The $Q$ samples can be randomly generated based the statistical distribution just at the beginning of the execution of the proposed scheme
for each time interval. 
Note that, whenever closed-form expressions or ones more concise than~(\ref{eq27_4}) are available
for $\mathbf{E}_{ \mathbb{H}_{csi}} \left[ \cdot \right]$, 
they should be utilized appropriately for efficient calculation~\cite{WangChinacom2018, r_JSAC_2021}.

The complexity of the proposed spectrum sharing scheme in Algorithm~\ref{SpectrumSharingScheme}
is mainly constituted by three parts.
They are the complexity of hierarchical clustering of the SU-satellite links in Algorithm~\ref{SUClusteringAlgorithm}, 
that of link scheduling for the CUs based on~(\ref{eq41_2}), 
and that of power control for the SUs based on~(\ref{eq23_1}) and~(\ref{eq53}).
\begin{itemize}
	\item
The complexity of Algorithm~\ref{SUClusteringAlgorithm} is mainly composed of 
calculation of the feature vectors $\mathcal{F}_{u}$ in~(\ref{eq35}),	
solving of the assignment problem in~(\ref{eq39_2}) for coarse link clustering, as well as fine link clustering for each coarse link cluster.
The feature vectors can be calculated with a complexity of $O(QMN_cN_s)$ where $M=I_{cl}F$.
The assignment problem in~(\ref{eq39_2}) can be solved with a complexity of $O(N_s^3)$~\cite{ref_AP}.
The complexity of fine link clustering can be expressed as $O(K\mathfrak{N}_sI_{cl}N_cN_s)$, 
where $\mathfrak{N}_s$ denotes the number of iterations corresponding to $\mathfrak{L}_s$~\cite{ref_K_means}.
\item
Based on complexity of the assignment problem in~(\ref{eq41_2}), i.e., $O(N_c^3)$,
link scheduling for the CUs served by all $M=I_{cl}F$ BSs has a complexity of $O(MN_c^3)$~\cite{ref_AP}.
\item
The complexity of power control for the SUs depends on
the number of subcarriers $K$, the number of iterations corresponding to $\mathfrak{L}_p$, denoted as $\mathfrak{N}_p$,
as well as complexity of the approximated convex problem of~(\ref{eq23_1}) in each iteration.
Taking the interior-point method as a benchmark, a rather conservative, or, pessimistic, 
evaluation for the complexity of each approximated problem of~(\ref{eq23_1}) is given by~(\ref{eq57}) in Appendix A.
Thus, the complexity of power control for the SUs can be 
written as $ O ( \sqrt{(MN_cN_s/K+2MN_c+2N_s)/K} \cdot \mathfrak{N}_p \cdot [Q(MN_c+N_s) +MN_cN_s/K+3MN_c+2N_s])$.
\end{itemize}
Simulations show that under the network settings in Section IV,
$\mathfrak{N}_s$ is always less than $15$, and $\mathfrak{N}_p$ is less than $10$.

Note that there is much potential for the complexity of power control to be further reduced.
For example, only the strongest SU-CU interference links are usually needed to be considered in practice~\cite{SatCom_integrate_book, SatCom_book, r_JSAC_2021}.
It indicates that most of the constraints in~(\ref{eq23_1}), especially those in~(\ref{eq23_1}b),
can be removed, which will undoubtedly lead to significant complexity reduction.
Besides, in cases when closed-form or more-concise expressions are available,
the complexity brought by $Q$ can also be largely reduced~\cite{WangChinacom2018, r_JSAC_2021}.
As the time scale for both satellite-terrestrial synchronization and spectrum sharing optimization,
the selection of $\Delta \tau$ provides another perspective for complexity adjusting. 
Basically, larger $\Delta \tau$ indicates less-frequent signaling exchange and computation,
and thus lower complexity.
Nevertheless, with larger $\Delta \tau$, the performance gain achieved by spectrum sharing tends to be smaller.
It is because with less-frequently-acquired CSI in $\mathbb{H}_{csi}$,
there may be more uncertainty of the channel fading state of the links 
due to the movement of SUs and CUs as well as the diversity of propagation environment.
This could cause undermatching between the SU-satellite and BS-CU links in spectrum sharing.
To balance performance and complexity,
systematic evaluation of the impact of the duration of $\Delta \tau$ is needed
for each specific network scenario in practice.

\section{Simulation Results and Discussion}
\subsection{Network Settings for Simulations}
\begin{figure}
	\centering
	\includegraphics[width=8.9cm]{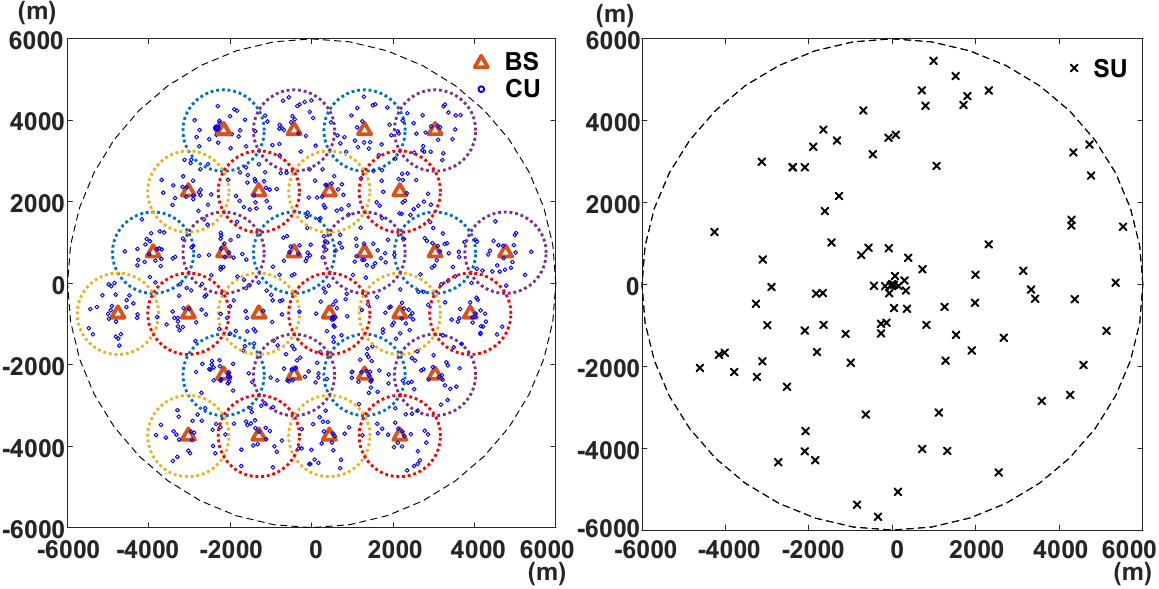}
	\caption{A random topology of the hybrid network for simulations.}
	\label{fig_simu_topology}
\end{figure}

The hybrid satellite-terrestrial network for simulations is shown in Fig.~\ref{fig_simu_topology}.
For clarity, it is shown in two separate parts.
Assume that $J=3$, $K=12$, $M=28$, $N_c = 24$, $N_s=96$, and $F=1$ or $4$.
Correspondingly, $I_{cl} = 28$ or $7$, $K' = 12$ or $3$, $N'_c = 2$ or $8$, and $N'_s = 8$.
Without loss of generality, it is assumed that $J=3$ satellites are all available for all the SUs, i.e., $J'_u = 3$.
The satellites are deployed in orbits at the altitude of $\mathcal{H}_{sat} = 500$km.
The radius for the coverage area of each BS is set as $1$km.
It is assumed that the CUs are randomly distributed within the coverage area of each BS,
and the SUs are randomly distributed in a circular area $\mathcal{A}$ slightly larger than total coverage area of all the BSs. 
Besides, the CUs $(i,r,v)$ and SUs $u$ are supposed to move along randomly-generated straight paths 
at random speeds of $\nu_{i,r,v}^{(cu,bs)} \in [0, 2]$m/s and $\nu_u^{(su)} \in [0, 10]$m/s, respectively.
Note that for clarity, only centers of the paths for the CUs and SUs are presented in Fig.~\ref{fig_simu_topology}.
Due to the relatively large distance between SUs and the satellites, 
a single location is considered for each satellite in simulations, sampled from its track across $T$.
The center of the circular area $\mathcal{A}$ is supposed to be located at $(116^{\circ} E, 40^{\circ} N)$,
and the sub-satellite points of the sampled locations for the satellites are 
at $(107^{\circ} E, 40^{\circ} N)$, $(116^{\circ} E, 40^{\circ} N)$, and $(125^{\circ} E, 40^{\circ} N)$, respectively.

The bandwidth of the subcarriers is set as $B=1$MHz, and the center frequency is $f_c = 2$GHz.
The receive antenna gains of the satellites are all set as  $|R_{u,j}^{(sat)}|^2 = 25$dBi, $\forall u, j$, and 
the transmit antenna gains of the BSs are $|T_{i,r,v}^{(bs)}|^2 = 15$dBi, $\forall k, i, v$.
The pattern of the transmit antenna array equipped at each SU is supposed to 
follow the ITU-R recommendation S.465
with a diameter of $0.5$m and main lobe width of $61.2^\circ$~\cite{10678835, ref_ITU_R_S465}.
Correspondingly, the transmit antenna gain of each SU for the SU-satellite link is $|T_{u}^{(su)}|^2 = 18.5$dBi,
and that for each SU-CU interference link, i.e., $|\hat{T}_{i,r,v,u}^{(cu,su)}|^2$,
determined by the off-axis angle of CU $(i,r,v)$ from the boresight of the transmit antenna array at SU $u$.
The transmit power of the BSs for each CU, i.e., $P_{bs}$, 
is supposed to be within $[0, 10]$dBm, and the maximum transmit power of the SUs is $P_{su} = 3$dBW.
The noise power is set as ${\sigma^{(cu)}}^2 = {\sigma^{(sat)}}^2 = \sigma^2 = -114$dBm.
The interference threshold is set as $\gamma_{th} = \sigma^2 \text{[dBm]} - 12.2$dB 
following the link protection criterion in the ITU-R Recommendation M.1799~\cite{ITURM1799},
which leads to a group of inter-link interference constraints for spectrum sharing.
The time scale $\Delta \tau$ is set as $10$s, and thus $T=\Delta \tau = 10$s.
Without loss of generality, 
set $\mathcal{C}_u^{QoS} = \min_{j=1,...,J'_u} \{ C^{(su)}_{u} |_{ p_{u}^{(su)}=P_{min}^{(su)}, \, \zeta_{u,j}=1} \} $
based on~(\ref{eq5_0}b) and~(\ref{eq8}), with $P_{min}^{(su)} = 10$dBm.

A general composite channel model consisting of large-scale and small-scale fading,
i.e., $h=\sqrt{\beta} \cdot w$, is adopted for all the BS-CU, SU-satellite, and SU-CU interference links~\cite{r_JSAC_2021, 10678835},
where $\beta$ denotes the large-scale fading, and $w$ is the small-scale fading.
$\beta$ is composed of path loss and shadowing, i.e., $\beta = l \cdot s$,
where $l$ is the path loss and $s$ represents the log-normal-distributed shadowing.
$w$ is supposed to follow Rayleigh distribution for the BS-CU and SU-CU interference links,
and Rician distribution for the SU-satellite links~\cite{10678835}.
$l$ for all the links is known as part of the statistical CSI in $\mathbb{H}_{csi}$.
Shadowing for the links is assumed to be $s[\text{dB}] = s_1[\text{dB}] + s_2[\text{dB}]$,
in which $s_1[\text{dB}]$ denotes the known part in $\mathbb{H}_{csi}$, generated based on a variance $\bar{\rho}$,
and $s_2[\text{dB}]$ is the random part with only the variance $\rho$ being known in $\mathbb{H}_{csi}$.
It is noted from~(\ref{eq5_0}) and~(\ref{eq6_0}) that the variance of shadowing is not needed for the SU-CU interference links.
Due to the fast varying of $w$, only the statistical parameter, i.e., the K-factor $\kappa$ for Rician fading, 
is supposed to be known in $\mathbb{H}_{csi}$.
Thus, the statistical CSI for the links in $\mathbb{H}_{csi}$, 
i.e., $\eta_{i,r,v}^{(cu,bs)}$, $\eta_{u}^{(sat,su)}$, and $\eta_{i,r,v,u}^{(cu,su)}$, could be expressed as 
$\{ l \cdot s_1,\, \rho \}$, $\{ l \cdot s_1,\, \rho, \kappa \}$, and $\{ l \cdot s_1\}$, respectively.

Based on the close-in (CI) free-space reference distance model~\cite{r_TVT_2016},
the path loss $l$ is expressed as $l [\text{dB}] = b + 10a\log_{10}(d/\bar{d}_0) +20\log_{10}(f_c)$,
where $d$ denotes transmission distance, $\bar{d}_0=1$m is the reference distance, $b$ is the path loss at $\bar{d}_0$,
and $a$ is the path loss exponent. 
$a$ and $b$ are set as $a^{(cu,bs)}=2.5$, $b^{(cu,bs)}=32.4$, $a^{(sat,su)}=2$, $b^{(sat,su)}=32.4$, 
and $a^{(cu,su)}=3$, $b^{(cu,su)}=32.4$, for the BS-CU, SU-satellite, and SU-CU links, respectively.
In the simulations, $l$ for the links is obtained based on the centers of the straight paths for the CUs and SUs.
As for the shadowing for the BS-CU and SU-satellite links, i.e., $s[\text{dB}] = s_1[\text{dB}] + s_2[\text{dB}]$, 
the variance of $s_1[\text{dB}]$ is set as $\bar{\rho} = 3$,
and a linear model is adopted for the variance of $s_2[\text{dB}]$,
i.e., $\rho$, according to the distance that CUs and SUs moved within $\Delta \tau$.
Specifically, $\rho$ is set as 
$\rho_{i,r,v}^{(cu,bs)} = \frac{\nu_{i,r,v}^{(cu)} \Delta \tau}{D_1^{ref}} \rho^{(cu,bs)}_{max} $ and 
$\rho_u^{(sat,su)} = \frac{\nu_u^{(su)} \Delta \tau}{D_2^{ref}} \rho^{(sat,su)}_{max} $
for the BS-CU and SU-satellite links, respectively.
It is assumed that $D_1^{ref} = \nu_{max}^{(cu)} \Delta \tau$, $D_2^{ref} = \nu_{max}^{(su)} \Delta \tau$,
$\nu_{max}^{(cu)} = 2$m/s, $\nu_{max}^{(su)} = 10$m/s, $\rho^{(cu,bs)}_{max} = 2$, and $\rho^{(sat,su)}_{max} = 2$.
The K-factor of Rician distribution for the SU-satellite links is set as $\kappa = 10$.
$Q=1000$ is adopted for the Monte Carlo approximation in~(\ref{eq27_4}).

\subsection{Performance of the Proposed Spectrum Sharing Scheme}
To evaluate performance of the proposed spectrum sharing scheme,
the achieved average sum rate of the hybrid network is calculated for both $F=4$ and $F=1$.
For comparison, three benchmark schemes, i.e., \emph{FineSync}, \emph{PartialPreScheme}, and \emph{RandScheme}, are considered for $F=4$. 
Two of them, i.e., \emph{FineSync}, and \emph{RandScheme}, are considered for $F=1$.
\emph{FineSync} is a spectrum sharing scheme with time-slot-based fine satellite-terrestrial synchronization,
also developed based on cooperative optimization of link scheduling for the CUs 
and joint link scheduling and power control for the SUs.
Specially, for implementation efficiency, it is carried out based on
Algorithm~\ref{SpectrumSharingScheme} by replacing $K$ subcarriers with $KN_s$ time-frequency RBs, with one SU being served in each RB.
\emph{PartialPreScheme} is a cooperative satellite-terrestrial link scheduling scheme 
obtained partly following our previous work in~\cite{10678835}.
It is only considered for $F=1$ as
the scheme proposed in~\cite{10678835} can't support partial frequency reuse in the terrestrial component.
In \emph{PartialPreScheme},
the SU-satellite links are scheduled based on link clustering following~\cite{10678835},
and feature vectors for the SU-satellite links are designed based on average transmission rate of the BS-CU links only, as
\begin{equation}\label{eq55}
\hat{\mathcal{F}}_{u} = \left[ \hat{\mathcal{F}}^{sub}_{u,1}, ..., \hat{\mathcal{F}}^{sub}_{u,F}  \right],
\end{equation}
with
\begin{equation}\label{eq56}
	\hat{\mathcal{F}}^{sub}_{u,r} = \left[ \hat{f}_{1,r,1}^{(u)}, ..., \hat{f}_{1,r,N_c}^{(u)}, \hat{f}_{2,r,1}^{(u)}, ..., \hat{f}_{2,r,N_c}^{(u)}, ...,  \hat{f}_{I_{cl},r,N_c}^{(u)}  \right],
\end{equation}
%$\hat{\mathcal{F}}_{u} = \left[ \hat{\mathcal{F}}^{sub}_{u,1}, ..., \hat{\mathcal{F}}^{sub}_{u,F}  \right]$, 
%with $\hat{\mathcal{F}}^{sub}_{u,r} = \left[ \hat{f}_{1,r,1}^{(u)}, ..., \hat{f}_{1,r,N_c}^{(u)}, \hat{f}_{2,r,1}^{(u)}, ..., \hat{f}_{2,r,N_c}^{(u)}, ...,  \hat{f}_{I_{cl},r,N_c}^{(u)}  \right]$
and $\hat{f}_{i,r,v}^{(u)} = \Delta C^{(cu)}_{i,r,v} (\underline{\gamma}_{i,r,v}^{(u)})$.
For fairness in comparison, 
the BS-CU link scheduling in Algorithm~\ref{SpectrumSharingScheme} is adopted in \emph{PartialPreScheme}
to maximize the average sum rate, instead of minimizing energy consumption of the hybrid network as in~\cite{10678835}.
\emph{RandScheme} is a random spectrum sharing scheme, 
in which the CUs and SUs utilize the subcarriers independently without power control.
In both \emph{PartialPreScheme} and \emph{RandScheme}, 
the satellite with the smallest SU-Satellite distance is selected for each SU,
and transmit power of the SUs is set as 
$\bar{p}_{u}^{(k^*_u,max)} = \min \{ \min_{(i,r,v) \in \mathbb{V}_{k^*_u}} \bar{p}_{u}^{(i,r,v)}, \, P_{su} \}$ 
based on~(\ref{eq31}b) and~(\ref{eq32_0}), 
with $k^*_u$ denoting the subcarrier that SU $u$ is served in.
Noted that the QoS of all CUs is ensured in \emph{PartialPreScheme} and \emph{RandScheme}.

\begin{figure} [t]
	\centering
	\includegraphics[width=8.8cm]{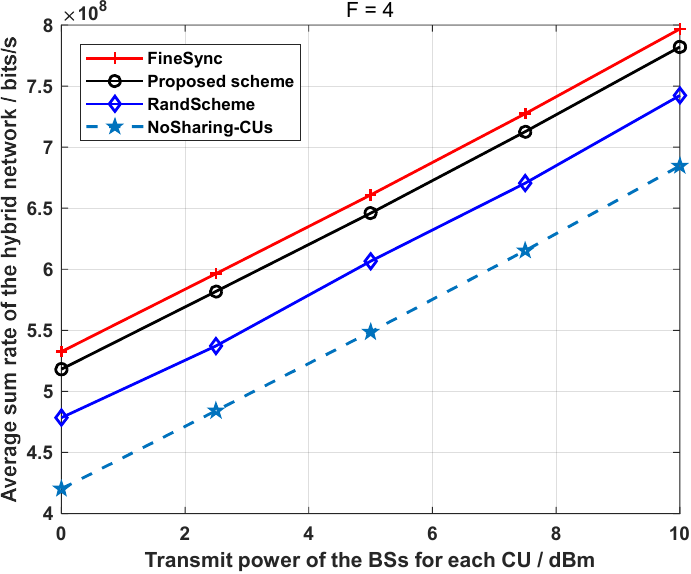}
	\caption{Average sum rate of the hybrid network for $F=4$.}
	\label{fig_SumRate}
\end{figure}

\begin{figure} [t]
	\centering
	\includegraphics[width=8.8cm]{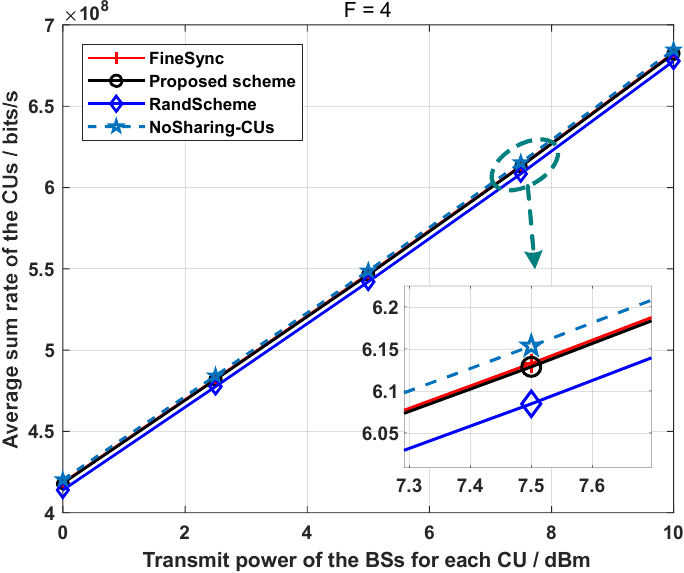}
	\caption{Average sum rate of the CUs achieved by different schemes for $F=4$.}
	\label{fig_SumRateCUs1}
\end{figure}

\subsubsection{\textbf{Performance evaluation for $F=4$}}
\ 

The average sum rate of the hybrid network achieved by the proposed scheme and the benchmark ones
for $F=4$ is shown in Fig.~\ref{fig_SumRate}.
For comparison, the average sum rate when only CUs are served without spectrum sharing by SUs
is also presented in Fig.~\ref{fig_SumRate}, denoted as \emph{NoSharing-CUs}.
Note that the results in Fig.~\ref{fig_SumRate} are obtained based on average among $10$ randomly-selected topologies of the hybrid network.
It can be seen that although strict inter-link interference constraints are adopted with $\gamma_{th} = \sigma^2 - 12.2$dB,
the proposed spectrum sharing scheme still achieves a significant improvement in average sum rate for the hybrid network.
Compared to \emph{NoSharing-CUs},
the improvement of average sum rate achieved by the proposed scheme is about $23\%$
when the transmit power of the BSs for each CU, i.e., $P_{bs}$, is $0$dBm,
and it remains at more than $15\%$ when $P_{bs}$ increases to $10$dBm.
Moreover, it can be observed that with only coarse satellite-terrestrial synchronization,
the proposed scheme achieves more than $85\%$ of the performance gain promised by the \emph{FineSync} scheme.
Besides, it can be observed from Fig.~\ref{fig_SumRateCUs1}, which shows the average sum rate of the CUs achieved by different schemes,
that the impact of inter-link interference on the CUs caused by the proposed spectrum sharing scheme is very small and could be ignored.

\begin{figure} [t]
	\centering
	\includegraphics[width=8.8cm]{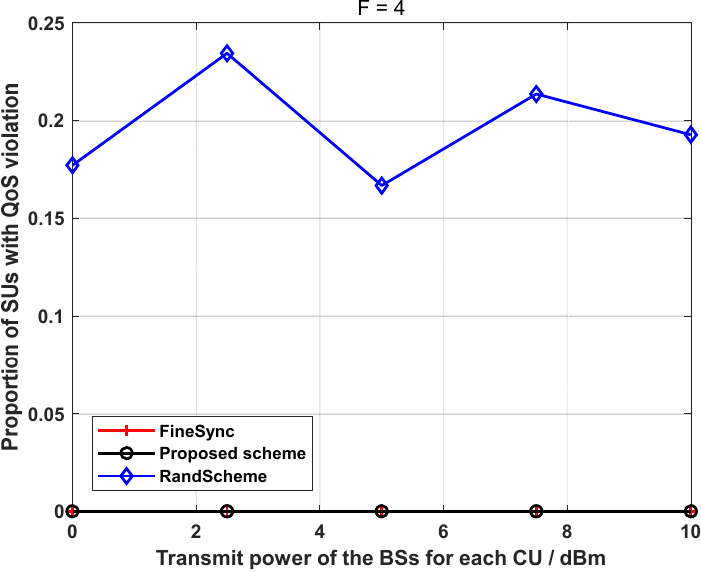}
	\caption{Proportion of SUs with QoS violation for $F=4$.}
	\label{fig_QoSviolation}
\end{figure}

\begin{figure} [t]
	\centering
	\includegraphics[width=8.8cm]{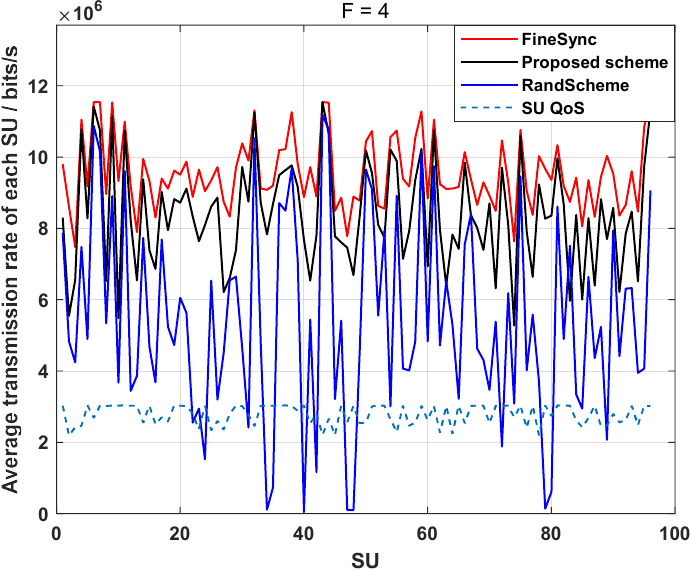}
	\caption{Average transmission rate of each SU in a random topology of the hybrid network for $F=4$.}
	\label{fig_Cap_SU}
\end{figure}

\begin{figure} [t]
	\centering
	\includegraphics[width=8.8cm]{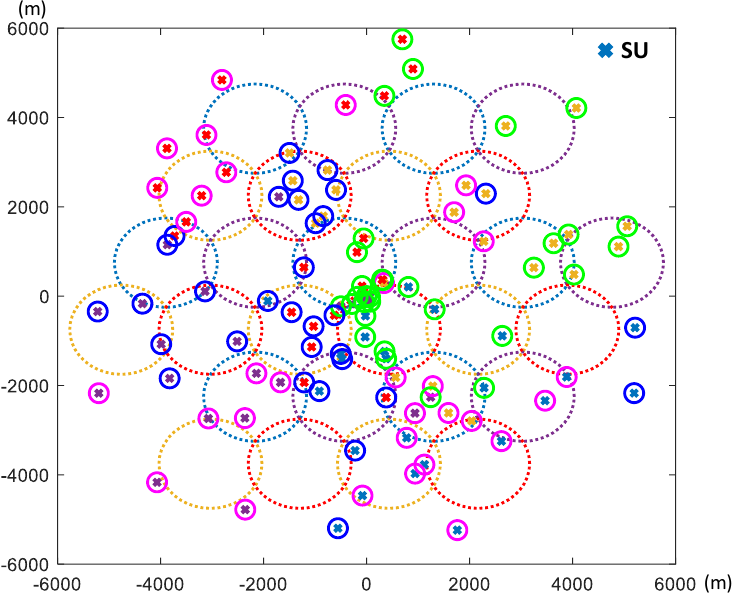}
	\caption{Scheduling result of the SU-satellite links for a random topology of the hybrid network obtained by the proposed scheme when $F=4$.}
	\label{clustering_whole}
\end{figure}

\begin{figure} [t]
	\centering
	\includegraphics[width=8.8cm]{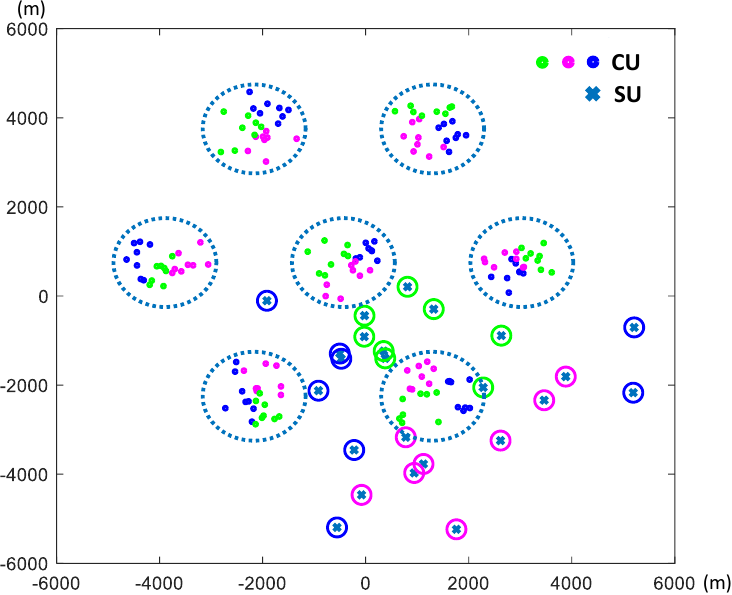}
	\caption{SU-satellite links and BS-CU links scheduled in subcarriers allocated to the BSs in FRC $r=1$ corresponding to Fig.~\ref{clustering_whole}.}
	\label{clustering_part}
\end{figure}

To further show the performance advantage of the proposed scheme, 
the QoS achieved for the SUs by different schemes when $F=4$ is presented in Fig.~\ref{fig_QoSviolation} and Fig.~\ref{fig_Cap_SU}.
Specifically, the proportion of SUs whose QoS requirements are violated under the strict inter-link interference constraints
is presented in Fig.~\ref{fig_QoSviolation},
and the average transmission rate of each SU in a random topology of the hybrid network is presented in Fig.~\ref{fig_Cap_SU}.
Although Fig.~\ref{fig_SumRate} indicates that an improvement of the average sum rate is also achieved by the \emph{RandScheme}, 
Fig.~\ref{fig_QoSviolation} and Fig.~\ref{fig_Cap_SU} 
demonstrate that QoS requirements of about $20\%$ of the SUs cannot be satisfied on average
in order to ensure QoS for the CUs in the \emph{RandScheme}.
On the contrary, QoS requirements of all the SUs are satisfied by both the proposed scheme and the \emph{FineSync} scheme.

To explain the rational behind the commendable performance of the proposed scheme,
Fig.~\ref{clustering_whole} and Fig.~\ref{clustering_part} show
the scheduling result of the SU-satellite links and BS-CU links in a random topology of the hybrid network
obtained by the proposed scheme when $F=4$.
In both Fig.~\ref{clustering_whole} and Fig.~\ref{clustering_part},
SU-satellite links scheduled in different subcarriers are indicated by colors of both cross markers
denoting the SUs and circles around the SUs.
Cross markers with the same color indicate SU-satellite links scheduled in subcarrers allocated to BSs in the same FRC.
Among each group of cross markers with the same color,
different colors of the circles around the cross markers further indicate the specific subcarrier 
that the corresponding SU-satellite links are scheduled in.
Similarly, different colors of the points denoting CUs in Fig.~\ref{clustering_part} indicate BS-CU links scheduled in different subcarriers.
SU-satellite links and BS-CU links with the same color are scheduled in the same subcarrier.
It can be observed from Fig.~\ref{clustering_whole} and Fig.~\ref{clustering_part}
that the spatial distribution of the CUs and SUs is well considered
in the link-feature-sketching-aided link scheduling in the proposed scheme via its impact on the statistical CSI of the links.
CUs and SUs scheduled in the same subcarrier are well separated with each other by relatively-large distances.

\begin{figure} [t]
	\centering
	\includegraphics[width=8.8cm]{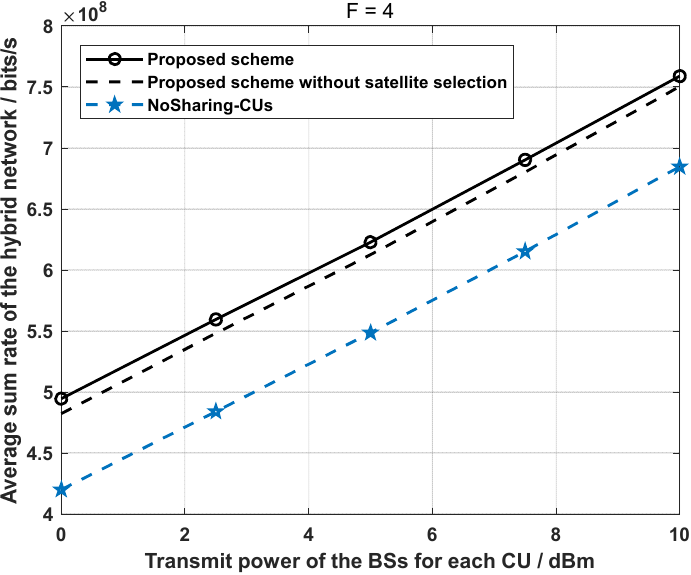}
	\caption{Average sum rate with and without satellite selection.}
	\label{fig_satellite_selection}
\end{figure}

\begin{figure} [t]
	\centering
	\includegraphics[width=8.8cm]{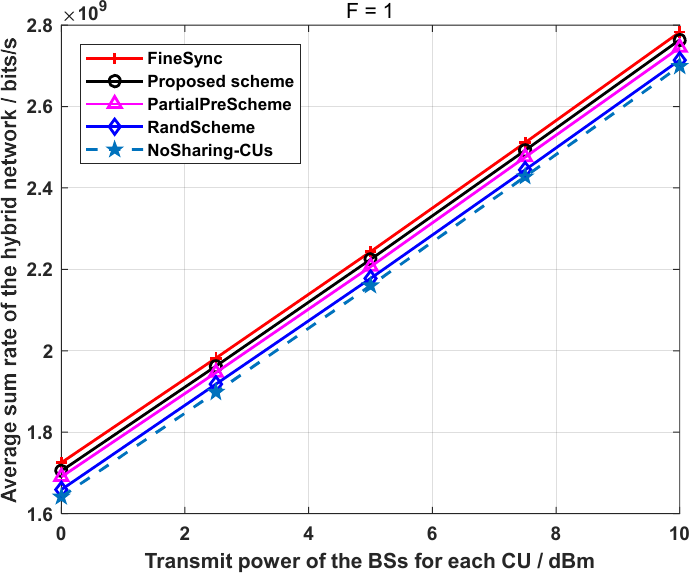}
	\caption{Average sum rate of the hybrid network for $F=1$.}
	\label{fig_SumRate2}
\end{figure}

It is noticed that among the $3$ satellites available for all SUs, 
only the one with its sub-satellite point at $(116^{\circ} E, 40^{\circ} N)$ is actually selected in the simulation results
for the proposed scheme.
That is owe to its overwhelming advantage in strength of the SU-satellite links brought by shorter distances to the SUs
than the other two satellites.
To further show the benefit of satellite selection, Fig.~\ref{fig_satellite_selection} presents 
the average sum rate achieved by the proposed scheme, both with and without satellite selection,
when only the other two satellites are available for the SUs. 
With their sub-satellite points being located at $(107^{\circ} E, 40^{\circ} N)$ and $(125^{\circ} E, 40^{\circ} N)$, 
the two satellites are similar in strength of the SU-satellite links for the SUs.
For the proposed scheme without satellite selection, each SU is randomly served by one of the two satellites.
The average sum rate without spectrum sharing is also presented for comparison.
It can be seen that a performance gap exists between the proposed scheme with and without satellite selection.
It shows the benefit of satellite selection.

\subsubsection{\textbf{Performance evaluation for $F=1$}}
\ 

\begin{figure} [t]
	\centering
	\includegraphics[width=8.8cm]{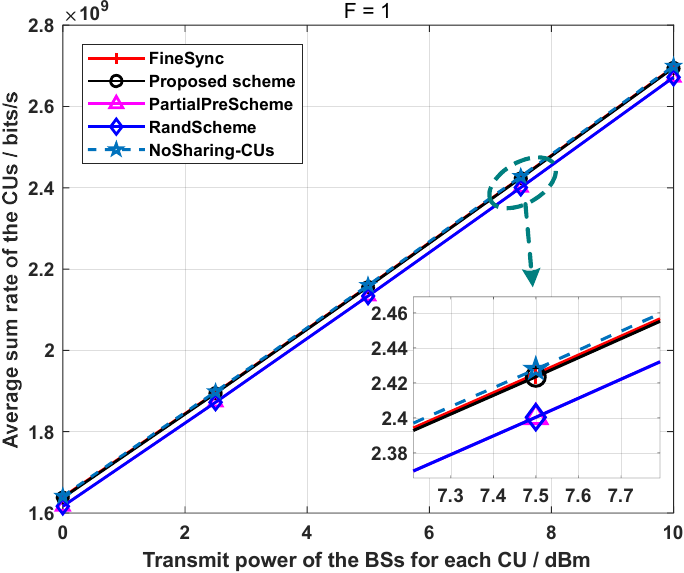}
	\caption{Average sum rate of the CUs achieved by different schemes for $F=1$.}
	\label{fig_SumRateCUs2}
\end{figure}

\begin{figure} [t]
	\centering
	\includegraphics[width=8.8cm]{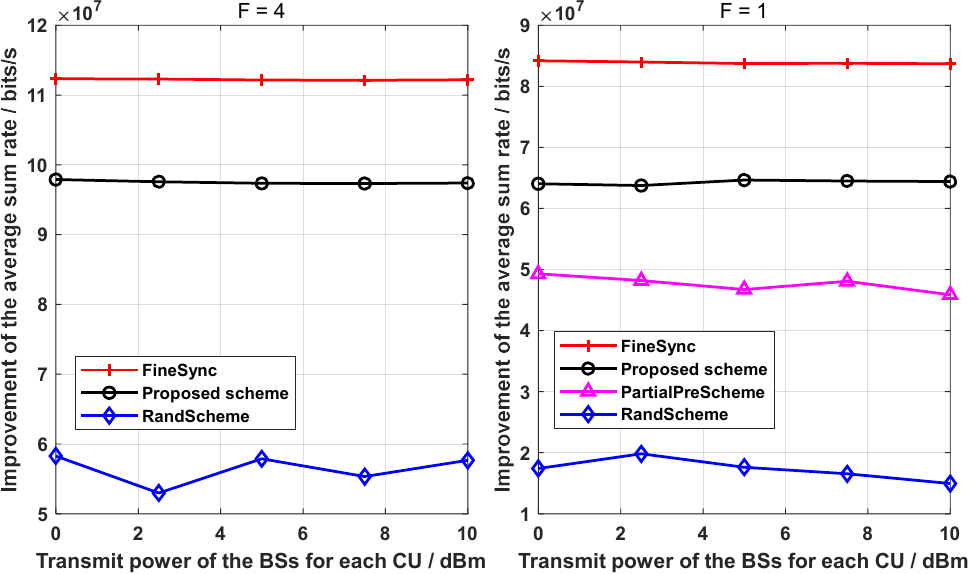}
	\caption{Improvement of the average sum rate of the hybrid network for $F=4$ and $F=1$.}
	\label{fig_RateImprovement}
\end{figure}

The average sum rate of the hybrid network with $F=1$, i.e., when full frequency reuse is adopted among the BSs, 
is shown in Fig.~\ref{fig_SumRate2}.
The results are also obtained based on average among $10$ randomly-selected topologies of the hybrid network.
It can be seen from Fig.~\ref{fig_SumRate2} that 
there is an obvious decrease in the performance gain achieved by the proposed scheme for $F=1$
compared to $F=4$ as shown in Fig.~\ref{fig_SumRate}.
The main reason for this is that the average sum rate of the CUs increases by about $300\%$ 
in Fig.~\ref{fig_SumRate2} compared to Fig.~\ref{fig_SumRate} due to the full frequency reuse among BS-CU links,
as indicated by Fig.~\ref{fig_SumRateCUs2} and Fig.~\ref{fig_SumRateCUs1}.
It straightforwardly leads to a decrease in the percentage of average sum rate improvement. 
The other reason is that when the frequency reuse factor $F$ for the BSs changes from $4$ to $1$,
a decrease in average distance from SUs to CUs sharing the same subcarriers, 
and further an increment of the overall strength of the SU-CU interference links, happen. 
Nevertheless, it can be observed from Fig.~\ref{fig_RateImprovement} that
most of the absolute improvement in the average sum rate of the hybrid network remains from $F=4$ to $F=1$.
With a negligible impact on the average transmission rate of the CUs as indicated by Fig.~\ref{fig_SumRateCUs2},
the extra transmission service provided to the SUs in the same frequency band 
is still rather attracting for practical applications.

%\begin{figure} [t]
%	\centering
%	\includegraphics[width=8.8cm]{SumRate2.pdf}
%	\caption{Average sum rate of the hybrid network for $F=1$.}
%	\label{fig_SumRate2}
%\end{figure}

\begin{figure} [t]
	\centering
	\includegraphics[width=8.8cm]{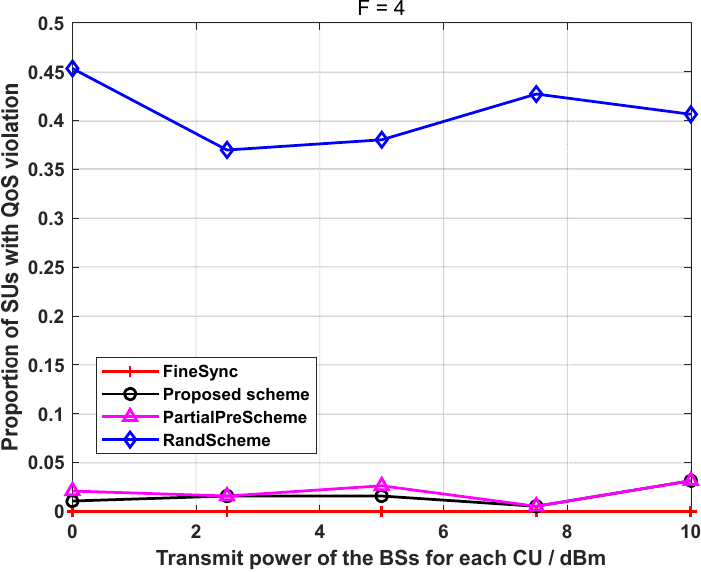}
	\caption{Proportion of SUs with QoS violation for $F=1$.}
	\label{fig_QoSviolation2}
\end{figure}

\begin{figure} [t]
	\centering
	\includegraphics[width=8.8cm]{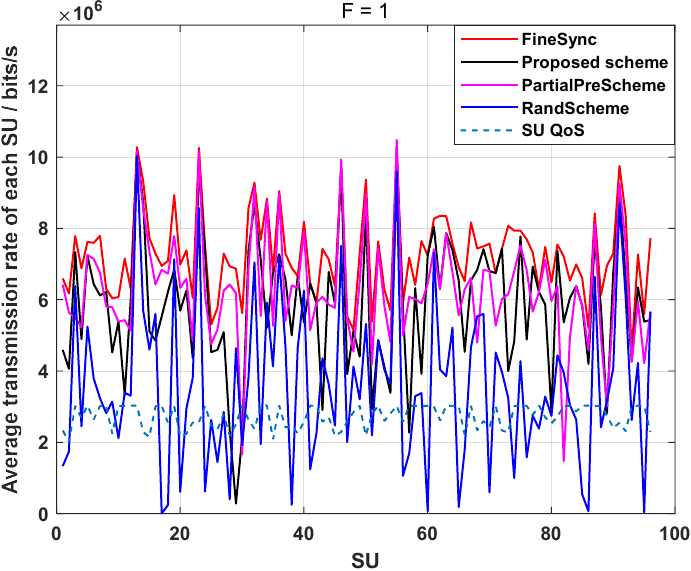}
	\caption{Average transmission rate of each SU in a random topology of the hybrid network for $F=1$.}
	\label{fig_Cap_SU2}
\end{figure}

Fig.~\ref{fig_QoSviolation2} and Fig.~\ref{fig_Cap_SU2} show the QoS achieved for the SUs for $F=1$.
The proportion of SUs whose QoS requirements are violated is presented in Fig.~\ref{fig_QoSviolation2},
and the average transmission rate of each SU in a random topology of the hybrid network is presented in Fig.~\ref{fig_Cap_SU2}.
It can be known from Fig.~\ref{fig_QoSviolation2} and Fig.~\ref{fig_QoSviolation} that 
to ensure QoS for the CUs,
the average percentage of SUs with QoS violation increases from about $20\%$ to more than $40\%$ as $F$ changes from $4$ to $1$
in the \emph{RandScheme}.
Contrarily, the percentage of SUs with QoS violation only increases to about $1.5\%$ on average
in the proposed scheme when $F=1$.
Note that the existence of QoS violation for a few SUs doesn't mean that 
the proposed scheme is infeasible for the hybrid network.
It can be avoided by keeping an appropriate margin for QoS requirements of the SUs and CUs, 
e.g., through loosening the assumption of full occupation of the same set of subcarriers by both SUs and CUs 
across the whole serving time. 
Reserving a performance margin is a safe routine in practice. 

Further, the performance advantage of the proposed scheme compared to our previous work in~\cite{10678835} can also be 
observed from Figs.~\ref{fig_SumRate2}--\ref{fig_QoSviolation2}.
Note that Fig.~\ref{fig_Cap_SU2} seems to indicate a superiority of the \emph{PartialPreScheme} over the proposed scheme
in terms of average transmission rate for the SUs,
as a significant amount of SUs are provided with larger average transmission rate in the \emph{PartialPreScheme}.
However, it is noticed from Fig.~\ref{fig_SumRateCUs2} that this happens
through a larger sacrifice of the average transmission rate for the CUs.

\section {Conclusions}
In this paper, statistical-CSI-based spectrum sharing has been explored 
for hybrid satellite-terrestrial networks with coarse network-wide time synchronization.
A time-scale-adaptable spectrum sharing framework has been proposed,
in which joint link scheduling and power control are implemented
for maximization of the average sum rate of the network while ensuring QoS for both CUs and SUs.
Satellite selection is supported to adapt to the case when multiple satellites are available in the satellite component.
For generality, 
both full and partial frequency reuse could be adopted among BSs in the terrestrial component.
With the framework,
a low-complexity spectrum sharing scheme has been proposed based on link-feature-sketching-aided hierarchical link clustering
and Monte-Carlo-and-successive-approximation-aided transmit power optimization.
Simulation results demonstrate that by link feature sketching, the spatial distribution of the CUs and SUs could be well utilized
for opportunistic spectrum sharing via its impact on statistical CSI of the links.
The proposed scheme promises a significant performance gain even under strict inter-link interference constraints.

\appendices
\section{Analysis on Complexity of the Approximated Convex Problem for Power Control}	
As presented in Lines 12--17 in Algorithm~\ref{SpectrumSharingScheme},
an approximated convex problem of~(\ref{eq23_1}) based on~(\ref{eq27_4}) and~(\ref{eq53}) 
needs to be solved in each iteration for power control. 
Taking the interior-point method with barrier functions for inequality constraints as a benchmark, 
the complexity of each approximated convex problem can be measured by
the number of basic searching steps, e.g., Newton steps, and the complexity of each step~\cite{ref_cvx_Boyd}.

Based on a rather conservative, or pessimistic, upper bound presented in~\cite{ref_cvx_Boyd},
the number of total Newton steps grows linearly with $\sqrt{MN_cN_s/K^2+2MN_c/K+2N_s/K}$,
where $MN_cN_s/K^2+2MN_c/K+2N_s/K$ is the number of inequality constraints in~(\ref{eq23_1}).
As for the complexity of each Newton step, it mainly lies in calculation of the 
gradient and Hessian of the combined target with barrier functions for the inequality constraints~\cite{ref_cvx_Boyd}.
It can be observed from~(\ref{eq23_1}), (\ref{eq27_4}), and~(\ref{eq53}) that 
the complexity of each Newton step increases linearly with $[Q(MN_c+N_s)+MN_cN_s/K+3MN_c+2N_s]/K$,
which denotes the number of basic terms of the optimization variables $p_{u}^{(su)}$ and $t_{i,r,v}$ 
in both the approximated target and inequality constraints of~(\ref{eq23_1}).

Thus, the complexity of each approximated convex problem of~(\ref{eq23_1}) based on~(\ref{eq27_4}) and~(\ref{eq53})
in Algorithm~\ref{SpectrumSharingScheme} can be conservatively, or pessimistically, expressed as 
\begin{equation}\label{eq57}
\begin{split}
	\!\!\!\! & O ( \sqrt{(MN_cN_s/K+2MN_c+2N_s)/K}   \\
	\!\!\!\! & \,\, \cdot [Q(MN_c+N_s) +MN_cN_s/K+3MN_c+2N_s]/K ).
\end{split}
\end{equation}

\end{spacing}
\end{document}